\documentclass[preprint,aps,amsmath,nofootinbib]{revtex4-1}
\usepackage{graphicx,array,dcolumn}
\usepackage{calc,tabularx, epsfig,mathrsfs}
\usepackage{hyperref}

\usepackage{amsmath,verbatim,enumerate}
\usepackage{amssymb}
\usepackage{wasysym}
\usepackage{xcolor}
\usepackage{multirow}
\usepackage{xspace}
\usepackage{slashed}
\usepackage{mathtools}
\usepackage{cancel}
\allowdisplaybreaks[1]
\newlength{\figurewidth}
\newcommand{\beq}{\begin{equation}}
\newcommand{\eeq}{\end{equation}}
\newcommand{\bea}{\begin{eqnarray}}
\newcommand{\eea}{\end{eqnarray}}
\newcommand{\ba}{\begin{array}}
\newcommand{\ea}{\end{array}}

\newcommand{\mn}{{\mu\nu}}

\newcommand{\pt}{\partial}

\newcommand{\al}{\alpha}
\newcommand{\bt}{\beta}

\newcommand{\ep}{\epsilon}
\newcommand{\ta}{\theta}
\newcommand{\lam}{\lambda}
\newcommand{\Lam}{\Lambda}
\newcommand{\G}{\Gamma}

\newcommand{\de}{\delta}

\newcommand{\OM}{\Omega}

\newcommand{\sg}{\sigma}

\makeatother

\begin{document}
%
\title{
Lorentzian Robin Universe of Gauss-Bonnet Gravity
}
\setlength{\figurewidth}{\columnwidth}
%
\author{Manishankar Ailiga}
\email{manishankara@iisc.ac.in}

\author{Shubhashis Mallik} 
\email{shubhashism@iisc.ac.in}

\author{Gaurav Narain}
\email{gnarain@iisc.ac.in}

\affiliation{
Center for High Energy Physics, Indian Institute of Science,
C V Raman Road, Bangalore 560012, India.
}

\vspace{5mm}

%
\begin{abstract}
The gravitational path-integral of Gauss-Bonnet gravity is investigated and the 
transition from one spacelike boundary configuration to another is analyzed. 
Of particular interest is the case of Neumann and Robin boundary conditions 
which is known to lead to a stable Universe in Einstein-Hilbert gravity
in four spacetime dimensions. After setting up the variational problem 
and computing the necessary boundary terms, the transition amplitude 
is computed \emph{exactly} in the mini-superspace approximation.
The $\hbar\to0$ limit brings out the dominant pieces in the path-integral 
which is traced to an initial configuration corresponding to 
Hartle-Hawking no-boundary Universe. A deeper study involving 
Picard-Lefschetz methods not only allow us to find the integration contour along 
which the path-integral becomes convergent but also aids in 
understanding the crossover from Euclidean to Lorentzian signature. Saddle analysis further 
highlights the boundary configurations giving dominant 
contribution to the path-integral which is seen to be 
those corresponding to Hartle-Hawking no-boundary proposal 
and agrees with the exact computation. To ensure completeness,
a comparison with the results from Wheeler-DeWitt equation is done. \\  \\ 
\text{KEYWORDS}: Lorentzian Path-Integral, Boundary Conditions, Gauss-Bonnet Gravity, Picard-Lefschetz Methods, No-Boundary Universe, Wheeler-DeWitt Equation.
\end{abstract}
\maketitle
\tableofcontents

\section{Introduction}
\label{intro}

In quantum field theory, path integral is a useful tool for computing 
the transition amplitude from one initial configuration to a final 
one. It allows us to obtain physical information about the 
system, most importantly scattering cross sections/S-matrix. 
In flat/curved spacetime quantum field theory, the path integral computation 
involves the integration over the fields defined on the background 
of fixed flat/curved spacetime, obeying appropriate boundary conditions. 

However, when gravity is present and/or background spacetime is dynamically 
evolving, as in the case of an expanding universe, defining a path-integral 
over fields is technically more involved. One has to 
carefully tackle issues like renormalizability, gauge-fixing, 
regularization, measure, boundary conditions and integration contour. 
Dealing with the last one comes with additional complications, which we aim to discuss here. 

The pure gravitational path-integral (ignoring matter sector) with 
metric as the only degree of freedom can be schematically written as
\bea 
\label{gp}
G[\rm Bd_i,Bd_f] && = \int_{\mathcal{M} 
+\partial \mathcal{M} } \mathcal{D}g_{\mu\nu} e^{i\rm S[g_{\mu\nu}]/\hbar}, 
\\
\label{EHact}
S[g_{\mu\nu}] && = \frac{1}{16\pi G_{\rm N}} \int d^Dx \sqrt{-g}
\left[-2\Lam + R + \cdots \right] + \text{boundary terms} \, 
\eea
where $g_{\mu\nu}$ is the metric, $S[g_{\mu\nu}]$ is the 
corresponding gravity action, $R$ is the Ricci scalar, 
$G_N$ is Newton's gravitational constant, 
$\Lam$ is the cosmological constant, 
$(\cdots)$ includes the higher derivative curvature terms. 
The path integral is defined on manifold $\mathcal{M}$ 
with boundary $\partial \mathcal{M} $. $\rm Bd_i$ and $\rm Bd_f$ 
refer to field configurations at the initial and final boundaries, respectively.

The gravitational path-integral mentioned above in eq. (\ref{gp}) is 
hard to define suitably where meaningful computations can 
be done. The biggest trouble is the lack of renormalizability. 
Even if one keeps the issue of renormalizability aside, 
one still has to deal with the issue of boundary terms and 
integration contour, which is quite non-trivial as the 
standard flat spacetime Wick-rotation no longer works. 
To understand the importance and subtle nature of the 
last two bits (boundary terms and integration contour), one can reduce the 
gravitational phase-space by considering mini-superspace approximation. 
This reduced phase space gives us more freedom to 
play around with the gravitational path-integral, thereby allowing 
us to learn more about the boundary conditions and 
the integration contours. 

In the mini-superspace approximation, the metric ansatz respecting 
spatially homogeneous and isotropic metric in $D$ spacetime 
dimensions is the FLRW metric. 
In polar
coordinates $\{t_p, r, \ta, \cdots \}$ it is given by
\beq
\label{eq:frwmet}
{\rm d}s^2 = - N_p^2(t_p) {\rm d} t_p^2 
+ a^2(t_p) \left[
\frac{{\rm d}r^2}{1-kr^2} + r^2 {\rm d} \OM_{D-2}^2
\right] \, .
\eeq
It has two unknown time-dependent functions: lapse $N_p(t_p)$
and scale-factor $a(t_p)$, $k=(0, \pm 1)$ is the scalar curvature, 
and ${\rm d}\OM_{D-2}$ is the 
metric for the unit sphere in $D-2$ spatial dimensions. 
In this reduced space, although
gravitational waves don’t exist,
a reduced version of diffeomorphism invariance still shows in the 
time co-ordinate $t_p$ and the dynamical scale-factor $a(t_p)$. 
This reduced setup is enough to explore the effects of 
boundary conditions and integration contour in the gravitational 
path-integral.

The path integral in this reduced gravitational phase space is
\bea
\label{eq:Gform_sch_fpt}
G[{\rm Bd}_f, {\rm Bd}_i]
&& = \int_{{\rm Bd}_i}^{{\rm Bd}_f} {\cal D} N_p {\cal D} \pi {\cal D} 
a(t_p) {\cal D} p {\cal D}  c {\cal D} \bar{c}  {\cal D} \rho {\cal D} \bar{\rho}
\notag\\
&&\quad \quad \times \exp \biggl[
\frac{i}{\hbar} \int_0^1 {\rm d} t_p \left(
N^\prime_p \pi + a^\prime p +c^\prime \rho + \bar{c}^{\prime} \bar{\rho} - N_p H \right)
\biggr] \, ,
\eea
where along with the integration over the scale-factor $a(t_p)$, lapse $N_p(t_p)$ 
the Fermionic ghost $ c$ and the anti-ghost $\bar{c}$, we also have an integration over 
their respective conjugate momenta given by
$p$, $\pi$, $\rho$, and $\bar{\rho}$ . It should be specified that the
$({}^\prime)$ here denotes derivative 
with respect to $t_p$, and without any loss of generality 
the time co-ordinate 
is chosen to range from $0\leq t_p \leq 1$. 
${\rm Bd}_i$ and ${\rm Bd}_f$ refers to the field configuration at 
initial ($t_p=0$) and final ($t_p=1$) boundaries respectively. 
The Hamiltonian constraint $H$ consists of two parts:
gravitational part and ghost
\beq
\label{eq:Htwo}
H = H_{\rm grav}[a, p] + H_{\rm gh} [N, \pi, c, \bar{c}, \rho, \bar{\rho}] \, ,
\eeq
where $H_{\rm grav}$ is the Hamiltonian for the gravitational action 
and the Batalin-Fradkin-Vilkovisky (BFV) \cite{Batalin:1977pb}
ghost Hamiltonian is denoted by $H_{\rm gh}$. BFV ghost generalizes the standard Fadeev-Popov ghost 
based on BRST symmetry. Unlike the gauge theories, in gravitational theories, the constraint algebra doesn't close, thereby requiring BFV quantization. In minisuperspace, where we have only one constraint (Hamiltonian Constraint), the algebra trivially closes, but still, BFV quantization is preferable.
Even though the phase space is quite reduced in the 
mini-superspace approximation, there is still 
time reparametrization invariance, which needs to be suitably 
gauge-fixed. A simple gauge choice is the ADM gauge
also known as proper-time gauge $N^\prime_p=0$. 
More details on the BFV quantization process and BFV ghost
can be found in the seminal works
\cite{Teitelboim:1981ua,Teitelboim:1983fk,Halliwell:1988wc}. 

A large part of the above gravitational path-integral 
mentioned in eq. (\ref{eq:Gform_sch_fpt})
can be performed exactly. This includes path-integral over the 
ghost-sector: $c, \bar{c}$ and $\rho, \bar{\rho}$; 
the conjugate momenta: $\pi$ and $p$. The path integral over the ghosts and the corresponding momenta leads to a lapse-independent constant term. Ignoring this constant term, the path-integral left behind  is 
\beq
\label{eq:Gform_sch}
G[{\rm Bd}_f, {\rm Bd}_i]
= \int_{\bf C} {\rm d} N_p
\int_{{\rm Bd}_i}^{{\rm Bd}_f} {\cal D} a(t_p) \,\,
e^{i S[a, N_p]/\hbar} \, .
\eeq
%

This path-integral can be interpreted as follows: 
$\int {\cal D} a(t_p) \,\, e^{i S[a, N_p]/\hbar}$
gives the transition amplitude for the
Universe to evolve from one boundary configuration to another
in the proper time $N_p$, where ${\bf C}$ refers to the contour of Lapse integration.
Depending on the choice of integration range ${\bf C}$ takes, eq. (\ref{eq:Gform_sch}) leads to different objects. The $0 < N_p <\infty$
choice leads to causal evolution from 
one boundary configuration ${\rm Bd}_i$ to another ${\rm Bd}_f$ 
as discussed in \cite{Teitelboim:1983fh}, where $a_0<a_1$ will imply expanding 
Universe while $a_0>a_1$ will imply contracting Universe. Whereas the other choice, $-\infty< N_p <\infty$, takes the gauge invariance into account.
For $-\infty < N_p <0$, $a_0<a_1$ 
will imply an expanding Universe while $a_0>a_1$ will imply 
contracting Universe where the subscript is with respect to $t_p$ coordinate.
The $0 < N_p <\infty$ leads to the (causal) propagator, while $-\infty< N_p <\infty$ leads to the
wavefunction of the universe, which is a solution to the Wheeler-DeWitt equation. 
In this discussion, we focus on the latter case, with the 
lapse integration range $-\infty < N_p <\infty$. 
This choice allows us to account for all the dominant saddles in the path integral.

The gravitational action that we will focus on in this review article is 
given by following (see \cite{Narain:2021bff, Narain:2022msz, Ailiga:2023wzl}
for earlier works investigating the path integral of such 
gravitational theories)
\bea
\label{eq:act}
S = \frac{1}{16\pi G_{\rm N}} \int {\rm d}^Dx \sqrt{-g}
\biggl[
-2\Lam + R + \al
\biggl( R_{\mu\nu\rho\sg} R^{\mu\nu\rho\sg} - 4 R_\mn R^\mn + R^2 \biggr)
\biggr] \, . 
\eea
This is the Gauss-Bonnet gravity action
where $G_{\rm N}$ is the Newton's gravitational constant, 
$\Lam$ is the cosmological constant term,  
$\al$ is the Gauss-Bonnet (GB) coupling and $D$ is spacetime dimensionality. 
The mass dimensions of various couplings are: 
$[G_N] = M^{2-D}$, $[\Lam] = M^2$ and $[\al] = M^{-2}$.

The gravitational action in eq. (\ref{eq:act}) falls in the class of lovelock gravity theories 
\cite{Lovelock:1971yv,Lovelock:1972vz,Lanczos:1938sf}, and
is a special type of higher-derivative gravity consisting 
of only two-time derivatives of metric-field.
This has the privilege of keeping the dynamical evolution equation 
of the metric field second order in time.
It should be pointed out that Gauss-Bonnet gravitational theories 
also arises in the
low-energy effective action of the heterotic string theory 
with $\al>0$ 
\cite{Cheung:2016wjt,Zwiebach:1985uq,Gross:1986mw,Metsaev:1987zx}.
Interestingly, this coupling also has received observational constraints
\cite{Chakravarti:2022zeq} coming from the 
study of the gravitational wave (GW) data of the event 
GW150914, which also offered the first observational 
confirmation of the area theorem \cite{Isi:2020tac}.

In this short review article, we will focus on the path integral given in the 
eq. (\ref{eq:Gform_sch}) for the gravitational action specified by
eq. (\ref{eq:act}), and discuss the effects of the Neumann/Robin boundary conditions (NBC/RBC)
on the transition amplitude
(see 
\cite{York:1986lje,Brown:1992bq,Krishnan:2016mcj,
Witten:2018lgb,Krishnan:2017bte} for the role played by 
boundary terms).
Imposing RBC at the initial boundary is like specifying 
a combination of scale factor and its conjugate momenta to be fixed.
A particular class of the same is ``linear'' combination of scale factor and its 
conjugate momenta, characterized by a single parameter. 
These RBCs interpolate between Dirichlet BC and Neumann BC.
{Past studies involving specifying DBCs at the initial boundary 
showed unsuppressed growth of fluctuations around the saddles 
leading to instability in the no-boundary proposal of the Universe
\cite{Feldbrugge:2017kzv,DiTucci:2018fdg,Lehners:2018eeo,Feldbrugge:2017fcc,Feldbrugge:2017mbc,Matsui:2024bfn}
(see \cite{Lehners:2023yrj}
for review on the no-boundary proposal).
These studies motivated us to investigate whether NBC or RBC can lead 
to stable behaviour of perturbations, also at the same time highlighting 
that Dirichlet BC is perhaps not the right boundary 
condition for gravity \cite{Krishnan:2016mcj,Krishnan:2017bte}.}

Explorations with the Neumann boundary condition (NBC) have shown
that gravitational perturbations are well-behaved in the 
no-boundary proposal of the Universe
\cite{DiTucci:2019bui,Narain:2021bff, Lehners:2021jmv, DiTucci:2020weq, Narain:2022msz, Mondal:2023cxx, Ailiga:2024nkz}. 
In the case of mini-superspace approximation with metric 
having $\mathbb{R}\times \mathbb{S}^3$ it is possible to 
do the computation of the path-integral exactly \cite{Narain:2022msz}.
These exact computations showed some interesting features 
where it was seen that our Universe undergoes signature changes 
from Euclidean to Lorentzian \cite{Narain:2022msz}.
Studies involving RBCs in Einstein-Hilbert gravity 
were done in \cite{DiTucci:2019dji,Mondal:2022gyp,DiTucci:2019bui}
under the WKB approximation using Picard-Lefschetz methods. 
In all these works it was noticed that perturbations were 
well-behaved. Exact computations imposing RBCs at the
initial boundary have been done in \cite{Ailiga:2023wzl}. 
Effects of Gauss-Bonnet addition have been thoroughly investigated
in \cite{Narain:2021bff, Narain:2022msz, Ailiga:2023wzl}. 

This review also focuses on the impact of having Gauss-Bonnet addition
in the gravitational action and the role it plays at the boundaries. 
This is crucial as the path-integral is sensitive to the 
boundary conditions. Retaining the consistency of the 
variational problem, the boundary condition is appropriately chosen.
This still doesn't guarantee that the solutions respecting these 
will be a stable configuration. But it is crucial to work those 
boundary conditions which lead to stable perturbations.   

Such path-integrals need to be analyzed appropriately in a 
systematic manner in the arena of complex analysis. Earlier studies involving complex analysis methods to understand 
Euclidean gravitational path integral was done in 
\cite{Hawking:1981gb,Hartle:1983ai}. 
Boundary effects studies in the Euclidean quantum cosmology
was done in 
\textit{tunnelling} proposal made in 
\cite{Vilenkin:1982de,Vilenkin:1983xq,Vilenkin:1984wp} 
and 
\textit{no-boundary} proposal made in \cite{Hawking:1981gb,Hartle:1983ai,Hawking:1983hj}.
Later works investigating the contour choice in the 
path-integral involving the usage of 
complex analysis methods were also done in 
\cite{Halliwell:1988ik,Halliwell:1989dy,Halliwell:1990qr}.
More recent work \cite{Kontsevich:2021dmb} on the 
`allowability' criterion proposes a simple method to determine 
the allowable complex metric on which quantum field theories can be consistently defined. 
It is, however, yet to be understood and established whether 
this criterion is necessary or sufficient (see also recent works 
\cite{Witten:2021nzp} and \cite{Lehners:2021mah}).
Our approach avoids the standard process of Wick rotation used in the Euclidean approach
and aims to directly tackle the Lorentzian gravitational 
path-integral using the Picard-Lefschetz (PL) method. 

Picard-Lefschetz (PL) methodology generalizes the notion of standard 
Wick-rotation by suitably picking a contour along which 
oscillatory Lorentzian integrals can be performed in a reliable manner. 
It provides a framework that considers all the saddles of the path-integral and 
systematically picks only the ones which eventually lead to 
convergent path-integral. The original integration contour 
gets appropriately deformed into the integrals over 
the steepest descent contours passing through such saddles. 
Such contours termed \textit{Lefschetz thimbles}
together form the generalized Wick-rotated contour.
In the Lorentzian quantum cosmology studies, these methods 
were utilized to understand the nature of the path integral 
\cite{Feldbrugge:2017kzv,Feldbrugge:2017fcc,Feldbrugge:2017mbc,
Vilenkin:2018dch, Vilenkin:2018oja, Rajeev:2021xit}
and the effects of the boundary conditions 
\cite{DiTucci:2019dji,DiTucci:2019bui,
Narain:2021bff,Ailiga:2024nkz,Lehners:2021jmv,Narain:2022msz, Ailiga:2023wzl}

The plan of this article is as follows: In section \ref{intro}, 
we present the introduction and discuss the motivation for studying the gravitational path 
integral. Section \ref{minisup} discusses the mini-superspace approximation 
and outlines the surface terms required for the consistent variational problem. 
In section \ref{trans}, we compute the \emph{exact} transition amplitude for 
Einstein-Gauss-Bonnet gravity with Neumann/Robin boundary conditions. We
analyze the semiclassical behaviour along with the significance of Gauss-Bonnet in the section \ref{hbar0}. 
In section \ref{inter}, we discuss the significance of the Robin boundary condition. 
A brief overview of the Picard-Lefschetz technique is provided 
in the section \ref{PL}, followed by its application in the section \ref{spa} to 
compute the saddle point results and verify their consistency with the 
results obtained from the exact analysis. Section \ref{WDW1} outlines 
the connection between the path integral and canonical approaches. 
Finally, the article concludes with a summary and outlook in Section \ref{Conc}.\\

\section{Mini-superspace action and Boundary Terms}
\label{minisup}

The conformally flat FLRW metric mentioned in eq. (\ref{eq:frwmet}) 
has vanishing Weyl-tensor $C_{\mu\nu\rho\sg} =0$. The non-zero entries of the 
Reimann tensor for the $d$-dimensional FLRW is given by
\cite{Deruelle:1989fj,Tangherlini:1963bw,Tangherlini:1986bw} 
\bea
\label{eq:riemann}
R_{0i0j} &=& - \left(\frac{a^{\prime\prime}}{a} - \frac{a^\prime N_p^\prime}{a N_p} \right) g_{ij} \, , 
\notag \\
R_{ijkl} &=& \left(\frac{k}{a^2} + \frac{a^{\prime2}}{N_p^2 a^2} \right)
\left(g_{ik} g_{jl} - g_{il} g_{jk} \right) \, ,
\eea
where $g_{ij}$ is the $D-1$-dimensional spatial part of the FLRW metric
and $({}^\prime)$ denotes derivative with respect to co-ordinate $t_p$.
The non-zero entries of the Ricci-tensor for this metric are 
\bea
\label{eq:Ricci-ten}
R_{00} &=& - (D-1) \left(\frac{a^{\prime\prime}}{a} - \frac{a^\prime N_p^\prime}{a N_p} \right)
\, , 
\notag \\
R_{ij} &=& \left[
\frac{(D-2) (k N_p^2 + a^{\prime2})}{N_p^2 a^2}
+ \frac{a^{\prime\prime} N_p - a^\prime N_p^\prime}{a N_p^3} 
\right] g_{ij} \, ,
\eea
while the Ricci-scalar for FLRW is given by
\beq
\label{eq:Ricci0}
R = 2(D-1) \left[\frac{a^{\prime\prime} N_p - a^\prime N_p^\prime}{a N_p^3} 
+ \frac{(D-2)(k N_p^2 + a^{\prime2})}{2N_p^2 a^2} \right]
\, .
\eeq
Interestingly, for conformal-flat (Weyl-flat) metrics, there is an identity relating the 
Riemann tensor with the Ricci-tensor and the Ricci scalar. This is given by,
\bea
\label{eq:Riem_exp}
R_{\mu\nu\rho\sg} = \frac{R_{\mu\rho} g_{\nu\sg} - R_{\mu\sg}g_{\nu\rho}
+ R_{\nu\sg} g_{\mu\rho} - R_{\nu\rho} g_{\mu\sg}}{D-2}
- \frac{R (g_{\mu\rho} g_{\nu\sg} - g_{\mu\sg} g_{\nu\rho})}{(D-1)(D-2)} \,.
\eea
By exploiting this identity one can express $R_{\mu\nu\rho\sg} R^{\mu\nu\rho\sg}$
in terms of $R_\mn R^\mn$ and $R^2$ as follows
\beq
\label{eq:Reim2_exp}
R_{\mu\nu\rho\sg} R^{\mu\nu\rho\sg}
= \frac{4}{D-2} R_\mn R^\mn - \frac{2 R^2}{(D-1)(D-2)} \, .
\eeq
This is a very useful relation when it is utilized 
in the GB-gravity action then, we get the following for 
case of Weyl-flat metrics
\bea
\label{eq:actGB}
\int {\rm d}^Dx \sqrt{-g} && \left(
R_{\mu\nu\rho\sg} R^{\mu\nu\rho\sg}  - 4 R_\mn R^\mn + R^2
\right)
\notag \\
&&
= \frac{D-3}{D-2} \int {\rm d}^Dx \sqrt{-g} \left(
- 4 R_\mn R^\mn + \frac{D R^2}{D-1}
\right) \, .
\eea
The gravitational action stated in eq. (\ref{eq:act})
for the FLRW metric mentioned in eq. (\ref{eq:frwmet}) is given by
\bea
\label{eq:midSact}
&&
S = \frac{V_{D-1}}{16 \pi G_N} \int {\rm d}t_p
\biggl[
\frac{a^{D-3}}{N_p^2} \biggl\{
(D-1)(D-2) k N_p^3 - 2 \Lam a^2 N_p^3 - 2 (D-1) a a^\prime N_p^\prime
\notag \\
&&
+ (D-1)(D-2) a^{\prime2} N_p + 2 (D-1) N_p a a^{\prime\prime}
\biggr\}
+ (D-1)(D-2)(D-3) \al\biggl\{
\frac{a^{D-5}(D-4)}{N_p^3} 
\notag\\
&&
\times (kN_p^2 + a^{\prime2})^2 
+ 4 a^{D-4}\frac{{\rm d}}{{\rm d}t_p} 
\left(
\frac{k a^\prime}{N_p} + \frac{a^{\prime 3}}{3N_p^3}
\right)
\biggr\}
\biggr] \, ,
\eea
where $V_{D-1}$ is the volume of $D-1$ dimensional sphere (for $\kappa=1$) and is given by \footnote{For $\kappa=(0,-1)$, the spatial slices are non-compact and $V_{D-1}$ wouldn't be finite. In these cases, we have to either use an appropriate regularization scheme to regulate the volume or take non-trivial topologies.},
\beq
\label{eq:volDm1}
V_{D-1} = \frac{\G(1/2)}{\G(D/2)} \left(\frac{\pi}{k}\right)^{(D-1)/2} \, .
\eeq
 This is the action for scale-factor $a(t_p)$ and lapse
$N_p(t_p)$ in $D$-dimensions. In $D=4$, something interesting happens
in the terms proportional to $\al$. This term becomes total time-derivative. 
This is expected from Gauss-Bonnet gravity which is topological in $D=4$. 
The mini-superspace 
gravitational action then becomes the following in $D=4$
\beq
\label{eq:mini_sup_d4}
S = \frac{V_{3}}{16 \pi G_N} \int {\rm d}t_p
\biggl[
6k a N_p - 2 \Lam a^3 N_p - \frac{6 a^2 a^\prime N_p^\prime}{N_p}
+ \frac{6 a a^{\prime 2}}{N_p} + \frac{6 a^{\prime\prime} a^2}{N_p}
+ 24 \al \frac{{\rm d}}{{\rm d}t_p} 
\left(
\frac{k a^\prime}{N_p} + \frac{a^{\prime 3}}{3N_p^3}
\right)
\biggr] \, .
\eeq
Lapse and scale-factor rescaling allows one to recast this action into 
a simpler form. This change of variables is given by
\beq
\label{eq:rescale}
N_p(t_p) {\rm d} t_p = \frac{N(t)}{a(t)} {\rm d} t \, ,
\hspace{5mm}
q(t) = a^2(t) \, .
\eeq
Then the original FLRW metric mentioned in eq. (\ref{eq:frwmet})
changes into the following
\beq
\label{eq:frwmet_changed}
{\rm d}s^2 = - \frac{N^2}{q(t)} {\rm d} t^2 
+ q(t) \left[
\frac{{\rm d}r^2}{1-kr^2} + r^2 {\rm d} \OM_{D-2}^2
\right] \, .
\eeq
Under this transformation, the gravitational action becomes the following 
\bea
\label{eq:Sact_frw_simp}
S = \frac{V_3}{16 \pi G_N} \int_0^1 {\rm d}t \biggl[
(6 k - 2\Lam q) N + \frac{3 \dot{q}^2}{2N}
+ 3q \frac{{\rm d}}{{\rm d} t} \left(\frac{\dot{q}}{N} \right)
+ 24 \al \frac{{\rm d}}{{\rm d} t} \left(
\frac{k\dot{q}}{2N} + \frac{\dot{q}^3}{24 N^3} 
\right)
\biggr] \, .
\eea
In this the $(\dot{})$ represent time $t$-derivative. 
For some terms, one can do the integration by parts which allows us to simplify a bit more 
and combine some of the terms. This eventually leads to
\bea
\label{eq:Sact_frw_simp_inp}
S &&
=\frac{V_3}{16 \pi G_N} \int_0^1 {\rm d}t \biggl[
(6 k - 2\Lam q) N - \frac{3 \dot{q}^2}{2N} \biggr] 
+ \frac{V_3}{16 \pi G} \biggl[
\frac{3q_f \dot{q}_f}{N} - \frac{3q_i \dot{q}_i}{N}
\notag \\
&&
+ 24 \al \left(
\frac{k\dot{q_f}}{2N} + \frac{\dot{q_f}^3}{24 N^3} 
- 
\frac{k\dot{q_i}}{2N} - \frac{\dot{q_i}^3}{24 N^3} 
\right)
\biggr]
\, .
\eea
Notice that there are two total derivative terms: one arises from the 
EH-part of gravitational action, while the other is the GB term itself which 
is topological in $4D$. We will keep $\Lambda$ to be arbitrary through the following sections till section \ref{hbar0}, where we take it to be positive and do saddle point studies. Note also that in the following, we will make use of 
the convention $V_3 = 8\pi G_N$. 

At this point, one can set up the variational problem by varying the action 
given in eq. (\ref{eq:Sact_frw_simp}) with respect to $q(t)$, which we do
it by working in the ADM gauge $\dot{N}=0$ (setting $N(t) = N_c$, a constant).
This can be implemented systematically by writing 
\beq
\label{eq:qfluc}
q(t) = \bar{q}(t) + \ep \de q(t)
\eeq
where one assumes $\bar{q}(t)$ to satisfy the equation of motion, 
$\de q(t)$ is the fluctuation around this. The parameter $\ep$ is inserted 
to keep track of terms in expansion, which becomes handy as one goes 
to higher orders in perturbative expansion. 
\beq
\label{eq:Sexp_qvar}
\de S = \frac{\ep}{2} \int_{0}^{1} {\rm d}t \biggl[
\left(-2 \Lam N_c + \frac{3 \ddot{\bar{q}}}{N_c} \right) \de q
+ \frac{3}{N_c} \frac{{\rm d}}{{\rm d} t} \left(\bar{q} \de \dot{q} \right)
+ 24 \al \frac{{\rm d}}{{\rm d} t} \left\{ 
\left(\frac{k}{2N_c} + \frac{\dot{q}^2}{8N_c^3} \right) \de \dot{q} \right\}
\biggr] \, .
\eeq
Note that in the above there are two total time-derivative terms. 
These eventually become boundary terms. 
For a consistent variational problem, these terms must either vanish 
for the given choice of boundary conditions or need to be 
appropriately cancelled by the addition of extra boundary terms. 
On the other hand, the collection of terms proportional to 
$\de q$ gives the equation of motion for $q$,
\beq
\label{eq:dyn_q_eq}
\ddot{\bar{q}} = \frac{2}{3} \Lam N_c^2 \, .
\eeq
The general solution to this equation is given by
\beq
\label{eq:qsol_gen}
\bar{q}(t) = \frac{\Lam N_c^2}{3} t^2 + c_1 t + c_2 \, .
\eeq
The unknown parameters $c_{1,2}$ gets fixed by satisfying the appropriate 
boundary conditions. The collection of total-derivative terms is given by
\beq
\label{eq:Sbd}
\delta S_{\rm bdy} = \frac{\ep}{2} \biggl[
\frac{3}{N_c} \left(\bar{q}_f \de \dot{q}_f - \bar{q}_i \de \dot{q}_i \right)
+ 24 \al \left\{ 
\left(\frac{k\de \dot{q}_f}{2N_c} + \frac{\dot{\bar{q}}_f^2\de \dot{q}_f}{8N_c^3} \right) 
-  \left(\frac{k\de \dot{q}_i}{2N_c} + \frac{\dot{\bar{q}}_i^2\de \dot{q}_i}{8N_c^3} \right)\right\}
\biggr] \, ,
\eeq
where 
\beq
\label{eq:BC_name}
\bar{q}_i = \bar{q}(t=0) \, , 
\hspace{5mm}
\bar{q}_f = \bar{q}(t=1) \, ,
\hspace{5mm}
\dot{\bar{q}}_i = \dot{\bar{q}}(t=0) \, ,
\hspace{5mm}
\dot{\bar{q}}_f = \dot{\bar{q}}(t=1) \, .
\eeq
The conjugate momenta corresponding to the field $q(t)$ can 
be computed from the action given in eq. (\ref{eq:Sact_frw_simp_inp})
\beq
\label{eq:mom_conju_qq}
\pi = \frac{\pt {\cal L}}{\pt \dot{q}} = - \frac{3\dot{q}}{2N_c} \, ,
\eeq
where we have used the ADM gauge. Utilizing the expression of the 
conjugate momentum, the boundary terms can be expressed in terms of 
$\pi$'s at the respective boundaries. This is given by
\beq
\label{eq:Sbd_mom}
\delta S_{\rm bdy} = - \ep\biggl[
\left(\bar{q}_f \de \pi_f - \bar{q}_i \de \pi_i \right)
+4 \al \left\{ 
\left(k\de \pi_f + \frac{\bar{\pi}_f^2\de \pi_f}{27} \right) 
-  \left(k\de \pi_i + \frac{\bar{\pi}_i^2\de\pi_i}{27} \right)\right\}
\biggr] \, .
\eeq
As mentioned before, for a consistent variational problem 
one has to get rid of these terms either by choosing boundary conditions 
where they vanish or by constructing suitable boundary action whose variation 
will cancel them exactly.

In the following, we will consider imposing Dirichlet boundary condition
at $t=1$, but at $t=0$ we will consider two possibilities: 
Neumann or Robin boundary 
conditions, as they lead to suppressed perturbations 
\cite{Feldbrugge:2017fcc,Feldbrugge:2017kzv}.

\subsection{Neumann Boundary condition (NBC) at $t=0$}
\label{neumann}

In this section, we impose Neumann boundary condition \cite{Krishnan:2016mcj,DiTucci:2019dji} 
at $t=0$. This means the initial momentum and the final position 
are fixed,
\beq
\label{eq:neuMa_cond}
\pi_i \,\, \&  \,\,
q_f = {\rm fixed} 
\hspace{5mm} 
\Rightarrow 
\hspace{5mm}
\de \pi_i = 0  \hspace{3mm} \&  \hspace{3mm}
\de q_f = 0 \, .
\eeq
Earlier studies have shown that this choice of boundary conditions 
lead to well-behaved perturbations which are suppressed
\cite{DiTucci:2019bui,Narain:2021bff,DiTucci:2020weq, Lehners:2021jmv}.
With the imposition of the above boundary condition, some of the 
terms in the eq. (\ref{eq:Sbd_mom}) will vanish. The residual terms 
need to be appropriately cancelled by addition of the surface terms 
whose variation will cancel them. The necessary surface terms 
are the following 
\beq
\label{eq:Sact_surf_NBC}
\biggl. S_{\rm surface} \biggr|_{\rm NBC}
= \frac{1}{2} \biggl[
-\frac{3q_f \dot{q}_f}{N_c} 
- 24 \al \left(
\frac{k\dot{q}_f}{2N_c} + \frac{\dot{q}_f^3}{24 N_c^3} 
\right) 
\biggr] 
= q_f \pi_f 
+ 4 \al \left(k\pi_f +\frac{\pi_f^3}{27} \right)  \, .
\eeq
It should be mentioned here that the first term is the 
Gibbon-Hawking term for the mini-superspace while the second term 
a Chern-Simon like term at the final boundary.
The constants $c_{1,2}$ appearing in the solution to 
the equation of motion eq. (\ref{eq:qsol_gen}) can be now determined
giving the solution $\bar{q}(t)$ to be
\beq
\label{eq:qsol_nbc}
\bar{q}(t) = \frac{\Lam N_c^2}{3} (t^2-1) - \frac{2 N_c \pi_i}{3} (t-1) + q_f \, ,
\eeq
From this one can immediately note that $\bar{q}_i$ is related to $q_f$,
\beq
\label{eq:q0_nbc}
\bar{q}_i = q_f + \frac{2 N_c \pi_i}{3} - \frac{\Lam N_c^2}{3} \, .
\eeq
This need not be true for off-shell configurations where $q_i$ could be 
anything. The surface terms when combined with the 
action in eq. (\ref{eq:Sact_frw_simp_inp}) leads to the 
total action of the system. This is given by
\beq
\label{eq:Sact_frw_simp_nbc}
S_{\rm tot}[q,N_c]
=\frac{1}{2} \int_0^1 {\rm d}t \biggl[
(6 k - 2\Lam q) N_c - \frac{3 \dot{q}^2}{2N_c} \biggr] 
+ q_i \pi_i
+ 4 \al \left(k\pi_i +\frac{\pi_i^3}{27} \right)
\, ,
\eeq
Furthermore, if one plugs the solution to equation of motion in the 
above while also using the expression for $\bar{q}_i$ from 
eq. (\ref{eq:q0_nbc}) one gets the total on-shell action 
\beq
\label{eq:stot_onsh_nbc}
S_{\rm tot}^{\rm on-shell}[\bar{q}, N_c] = \frac{\Lam^2}{9} N_c^3 
- \frac{\Lam \pi_i}{3} N_c^2 
+ \left(3k - \Lam q_f + \frac{\pi_i^2}{3} \right) N_c
+ q_f \pi_i + 4\al \left(k\pi_i + \frac{\pi_i^3}{27} \right) \, .
\eeq
This is also the action for the lapse $N_c$. It is important to note
at this point that compared to the action obtained in the case of 
Dirichlet boundary condition, this doesn't have singularity 
at $N_c=0$ \cite{Feldbrugge:2017kzv,Feldbrugge:2017fcc,Feldbrugge:2017mbc,
DiTucci:2019bui,Narain:2021bff,
Lehners:2021jmv, DiTucci:2020weq}. 
The absence of singularity can be understood by realizing 
that in NBC path-integral as one doesn't fix $q_i$ at $t=0$, 
transitions from all possible values of $q_i$ to a fixed 
$q_f$ are allowed, including the instantaneous transition occurring with $N_c=0$. 
This is only true for the Neumann BC and need not hold 
once the boundary conditions get generalized to Robin BC,
in the sense that there will be a value of $N_c$ for which the action
will be singular.

\subsection{Robin Boundary condition (RBC) at $t=0$}
\label{robin_surf}

Let's now consider the case of Robin BC being imposed at the initial boundary
with Dirichlet BC at the final boundary.
This is a well posed boundary value problem where we fix 
the linear combination of $q_i$ and $\pi_i$ at the initial boundary, 
while keeping $q_f$ at the final boundary fixed,
\beq
\label{eq:robin_cond}
\pi_i + \bt q_i = P_i= {\rm fixed} \hspace{3mm} \&  \hspace{3mm}
q_f = {\rm fixed} 
\hspace{5mm} 
\Rightarrow 
\hspace{5mm}
\de P_i = 0  \hspace{3mm} \&  \hspace{3mm}
\de q_f = 0 \, .
\eeq
When we have these boundary conditions, then the boundary terms 
mentioned in eq. (\ref{eq:Sbd_mom}) reduces to the following 
\beq
\label{eq:Sbd_mom_RBC}
\biggl. \delta S_{\rm bdy} \biggr|_{\rm RBC} = - \ep\biggl[
\bar{q}_f \de \pi_f + \bt \bar{q}_i \de q_i 
+4 \al \left\{ 
k\de \pi_f + \frac{\bar{\pi}_f^2\de \pi_f}{9} 
+ \bt\left(k\de q_i + \frac{(\bar{P}_i- \bt \bar{q}_i)^2\de q_i}{9} \right)\right\}
\biggr] \, .
\eeq
As before for consistent variational problem, we require that these terms 
either vanish or should be appropriately cancelled by the addition 
of suitable surface terms. It is seen that if one adds the following terms to the boundary 
\bea
\label{eq:Sact_surf_RBC}
\biggl. S_{\rm surface} \biggr|_{\rm RBC} 
&=& q_f \pi_f 
+ 4 \al \left(k\pi_f +\frac{\pi_f^3}{27} \right) 
+ \frac{\bt}{2} (q_f^2 + q_i^2)
\notag \\
&&
+ 4 \al \bt \biggl[
k q_i + \frac{q_i}{9} \left(
P_i^2 - \bt P_i q_i + \frac{\bt^2 q_i^2}{3} 
\right)
\biggr]\, ,
\eea
it leads to a consistent variational problem \cite{Ailiga:2023wzl}.
The first two terms here are the same as those obtained in the case 
of Neumann BC, while the last two terms appear additionally in the 
case of Robin BC. Moreover, the term in the last line arises when Gauss-Bonnet 
contribution is taken into account and was obtained for the first time in 
\cite{Ailiga:2023wzl}. The additional terms are proportional to $\bt$ and 
vanish when $\bt\to0$ (Neumann limit). The total action after combining these 
surface terms with the action in eq. (\ref{eq:Sact_frw_simp_inp})
becomes
\beq
\label{eq:Sact_frw_simp_RBC}
S_{\rm tot}[q,N_c]
=\frac{1}{2} \int_0^1 {\rm d}t \biggl[
(6 k - 2\Lam q) N_c - \frac{3 \dot{q}^2}{2N_c} \biggr] 
+ q_i P_i + \frac{\bt}{2} (q_f^2 - q_i^2)
+ 4 \al \left(k P_i +\frac{P_i^3}{27} \right)
\, .
\eeq
Once again in the limit $\bt\to0$ this reduces to the total action for 
Neumann BC mentioned in eq. (\ref{eq:Sact_frw_simp_nbc})
as $P_i \to \pi_i$. 

The constants appearing in the solution of the equation of motion 
$c_{1,2}$ can now be determined. This gives the solution 
\beq
\label{eq:qsol_RBC}
\bar{q}(t) = \frac{\Lam N_c^2}{3} t^2 
+ \frac{P_i}{\bt}
+ \left(1 + \frac{3}{2 \bt N_c} \right)^{-1}
 \left(t + \frac{3}{2 \bt N_c} \right)
 \left(q_f - \frac{P_i}{\bt} - \frac{\Lam N_c^2}{3} \right) \, ,
\eeq
Once again we note that on putting $t=0$ in the above 
we get a relation between $q_i$ and $q_f$
\beq
\label{eq:q0_rbc}
\bar{q}_i =\frac{P_i}{\bt}
+\left(\frac{3}{3+2 \bt N_c} \right)
\left(q_f - \frac{P_i}{\bt} - \frac{\Lam N_c^2}{3} \right) \,.
\eeq
Off-shell $q_i$ can be anything. The on-shell total can now be worked out 
by making use of $\bar{q}(t)$ mentioned in eq. (\ref{eq:qsol_RBC})
into the action for RBC given in eq. (\ref{eq:Sact_frw_simp_RBC}). 
This is given by
\bea
\label{eq:stot_onsh_rbc}
S_{\rm tot}^{\rm on-shell}[\bar{q}, N_c] &=& \frac{1}{18(3+ 2 N_c \bt)} \biggl[
\bt \Lam^2 N_c^4 + 6 \Lam^2 N_c^3 +
N_c^2 \{108 \bt  k-18 \Lam  (P_i+\bt q_f)\}
\notag \\
&&
+18 N_c \left\{9
k+P_i^2+ q_f \left(\bt^2 q_f-3 \Lambda \right)\right\}
+ 54 P_i q_f
\biggr] 
+ 4 \al \left(k P_i +\frac{P_i^3}{27} \right) .
\eea
This is the lapse $N_c$ action in the RBC case.
Note that in the limit $\bt\to0$ this reduces to the 
lapse action for the NBC mentioned in 
eq. (\ref{eq:stot_onsh_nbc}).

\section{Transition Amplitude}
\label{trans}

Having computed the total action  for the lapse $N_c$, 
we proceed further to compute the transition amplitude from one 
three-dimensional spacelike hypersurface to another
(for both NBC and RBC cases). The quantity of interest here 
(in the mini-superspace approximation)
is as follows (see \cite{Halliwell:1988ik,Feldbrugge:2017kzv})
\beq
\label{eq:Gamp}
G[ {\rm Bd}_f, {\rm Bd}_i]
= \int_{\bf C}  {\rm d} N_c  
\int_{{\rm Bd}_i}^{ {\rm Bd}_f} {\cal D} q(t) \,\, 
\exp \left(\frac{i}{\hbar} S_{\rm tot}[q, N_c] \right)
\, , 
\eeq
where as before ${\rm Bd}_i$ and  ${\rm Bd}_f$ refers to initial and final 
boundary configurations respectively. 
The $S_{\rm tot}$ appearing above is the total action 
in the mini-superspace approximation. 
For the NBC case it is given in eq. (\ref{eq:Sact_frw_simp_nbc})
and for the RBC it is given in eq. (\ref{eq:Sact_frw_simp_RBC}).
The NBC path-integral is not too hard to compute 
\cite{Narain:2022msz, Ailiga:2023wzl}. The RBC path-integral 
is a bit non-trivial, but can be computed exactly by making 
use of the procedure mentioned in section 2 of \cite{Ailiga:2023wzl}. 

The relevant expressions that will be needed in the following are 
the path-integral of a one-particle moving in linear potential 
whose action is given by
\beq
\label{act1part1}
S_{\rm tot}[q] = S[q] + S_{\rm bd} = 
\int_0^1 {\rm d}t \left[ \frac{m}{2} \dot{q}^2 - \lam q \right]
+ S_{\rm bd} \, ,
\eeq
$m$ is a $t$-independent parameter and $\lam$ is coupling parameter,
$S_{\rm bd}$ is the surface term added to have a 
consistent variational problem. The path-integral we want to compute is  
\beq
\label{1partPI}
\bar{G}[{\rm Bd_f}, {\rm Bd_i}] =
\int_{\rm Bd_i}^{\rm Bd_f} {\cal D} q(t) \,\, e^{i S_{\rm tot}[q]/\hbar} \, .
\eeq
For the initial Neumann and Robin boundary conditions they 
are given by the following expressions, respectively
\bea
\label{Gnbc_v}
\bar{G}_{\rm NBC}(q_{f},t=1; p_{i},t=0)  &=& 
\exp \biggl[
\frac{i}{\hbar} \left\{
p_{i}q_{f}- \frac{p_{i}^2}{2m} 
- \frac{\lam \left(\lam - 3 p_{i} + 6m q_{f}\right)}{6m}
\right\}
\biggr]  \, ,
\\
\label{eq:Grbc_Gnbc}
\bar{G}_{\rm RBC} (Q_f, t=1; P_i, t=0)
&=& \sqrt{\frac{e^{i\bt Q_f^2}}{i \bt} }
\int_{-\infty}^{\infty} \frac{{\rm d} \tilde{p}}{\sqrt{2 \pi \hbar}}
e^{i\frac{(P_i - \tilde{p})^2}{2\bt\hbar}} \bar{G}_{\rm NBC}(q_f, t=1; \tilde{p}, t=0) \, ,
\eea
where $p_i$ is the initial momentum, $q_i$ in initial position, 
$P_i = p_{i}+\bt \,q_{i}$ and $Q_f = q_f$ is the final position. 
The details of this derivation 
can be found in section 2 of \cite{Ailiga:2023wzl}.

\subsection{NBC at $t=0$}
\label{NBCt0}

Let's consider the mini-superspace path-integral 
mentioned in eq. (\ref{eq:Gamp}) for the Neumann boundary condition. 
To compute this we make use of the expression given in 
eq. (\ref{Gnbc_v}). In order to utilize this, we first compare 
the actions in eq. (\ref{act1part1}) with the NBC mini-superspace action  
given in eq. (\ref{eq:Sact_frw_simp_nbc}). 
\begin{gather}
\label{eq:subsNBC}
m \to -\frac{3}{2 N_c} \, , 
\hspace{1cm}
V(q) = \lam q \to \Lam N_c q 
\hspace{3mm} \Rightarrow \hspace{3mm} \lam \to \Lam N_c \, ,
\hspace{1cm}
p_i \to \pi_i \, .
\end{gather}
This gives the following expression for the 
Neumann boundary condition
\bea
\label{eq:gbar_NBC_ms}
&&
\int_{{\rm Bd}_i}^{{\rm Bd}_f} {\cal D} q(t) 
\exp \left(\frac{i}{\hbar} S_{\rm tot}[q, N_c] \right)
= \exp\left(\frac{i}{\hbar} S_{\rm tot}^{\rm on-shell}[\bar{q}, N_c] \right) 
\notag \\
&=& \exp\left[\frac{i}{\hbar} \left\{3k N_c  
+ 4 \al \left(k \pi_i + \frac{\pi_i^3}{27}\right) \right\}\right]
\bar{G}_{\rm NBC}[q_f, t=1; \pi_i, t=0] \, ,
\eea
where $S_{\rm tot}^{\rm on-shell}[\bar{q}, N_c]$ is given in eq. (\ref{eq:stot_onsh_nbc})
and $\bar{G}_{\rm NBC}[q_f, t=1; \pi_i, t=0]$ is given by
\beq
\label{eq:GbarNBC_ms}
\bar{G}_{\rm NBC}[q_f, t=1; \pi_i, t=0]
= \exp \biggl[
\frac{i}{\hbar}\biggl\{
\frac{\Lam^2 N_c^3}{9} - \frac{\Lam \pi_i N_c^2}{3} 
+ \left(\frac{\pi_i^2}{3} - \Lam q_f\right) N_c + \pi_i q_f
\biggr\}
\biggr] \, .
\eeq
It should be emphasized that the path-integral can also be performed 
using the zeta-function \cite{Narain:2022msz}. However, the results differ from above by a 
numerical pre-factor. It is worth noting that the quantum fluctuation determinant for the Neumann boundary condition results in a constant, which is 1, that agrees with the Van Vleck-Pauli-Morette determinant (see \cite{doi:10.1142/7305}). For a more detailed analysis, see Appendix \ref{Append A}.

Note that $S_{\rm tot}^{\rm on-shell}[\bar{q}, N_c]$ doesn't posses a singularity
at $N_c=0$ and the path-integral over $q(t)$ doesn't give rise to 
additional singularities or branch-cuts. This is a feature of imposing 
Neumann BC at the initial boundary which doesn't hold for a 
generic boundary condition. This allows to push the 
limit of $N_c$-integration all the way up to $-\infty$ 
thereby giving
\beq
\label{eq:Gab_afterQ}
G_{\rm NBC}[{\rm Bd}_f, {\rm Bd}_i]
=  \int_{-\infty}^\infty {\rm d} N_c \,\, 
\exp \left(\frac{i}{\hbar} S_{\rm tot}^{\rm on-shell}[\bar{q}, N_c] \right) \, ,
\eeq
where $S^{\rm on-shell}_{\rm tot}[\bar{q}, N_c]$ is given in eq. (\ref{eq:stot_onsh_nbc}). By extending the range of integration all the way to $-\infty$, one constructs the `` wave function ", which is the solution of the WDW equation, as we later show in sec.\ref{WDW1}. This is in contrast to the equation in \ref{eq:Gamp}, which computes the ``propagator". 
The integral can be done exactly by doing a change of variables achieved by shifting the 
lapse $N_c$ by a constant 
\beq
\label{eq:NcTONb}
N_c = \bar{N} + \frac{\pi_i}{\Lam} \, 
\hspace{5mm}
\Rightarrow
\hspace{5mm}
{\rm d}N_c \hspace{3mm} \to \hspace{3mm} {\rm d} \bar{N} \, .
\eeq
Due to the change of variables the transition amplitude is given by
\bea
\label{eq:Gab_Nb}
G[{\rm Bd}_f, {\rm Bd}_i]
= &&
\frac{1}{2}\exp \left[\frac{i}{\hbar} 
\left(\frac{3}{\Lam} + 4\al \right) \left(k \pi_i + \frac{\pi_i^3}{27} \right) \right]
\notag \\
&& \times
\int_{-\infty}^\infty {\rm d} \bar{N} \,\, 
\exp \left[\frac{i}{\hbar} \left\{\frac{\Lam^2}{9} \bar{N}^3 + (3k - \Lam q_f) \bar{N} \right\} \right] 
= \Psi_1(\pi_i) \Psi_2(q_f)
\, ,
\eea
where 
\bea
\label{eq:psi1}
&&
\Psi_1(\pi_i) =  \exp \left[\frac{i}{\hbar} 
\left(\frac{3}{\Lam} + 4\al \right) \left(k \pi_i + \frac{\pi_i^3}{27} \right) \right] \, ,
\\
\label{eq:psi2}
&&
\Psi_2(q_f) = \int_{-\infty}^\infty {\rm d} \bar{N} \,\, 
\exp \left[\frac{i}{\hbar} \left\{\frac{\Lam^2}{9} \bar{N}^3 + (3k - \Lam q_f) \bar{N} \right\} \right] \, .
\eea
It is seen that the transition amplitude with Neumann BC at the initial time 
and Dirichlet BC at the final time is a product of two parts: $\Psi_1(\pi_i)$ and $\Psi_2(q_f)$.
The former is only dependent on the initial momentum, while the 
later is only a function of the final size of the Universe. This factorization has been 
observed in earlier studies of the Wheeler-DeWitt (WdW) equation in mini-superspace 
approximation of Einstein-Hilbert gravity
\cite{Lehners:2021jmv} (and also in \cite{Narain:2022msz}).

The function $\Psi_2(q_f)$ can be computed as discussed in detail in the paper 
\cite{Narain:2022msz}. It is proportional to Airy-function. This allow
us to write a closed form expression for the transition amplitude 
in the NBC case as 
\bea
\label{eq:Gbd0bd1_full_nbc}
G_{\rm NBC}[{\rm Bd}_f, {\rm Bd}_i]
= &&
\sqrt{3} \left(\frac{3\hbar}{\Lam^2}\right)^{\frac{1}{3}}
\exp \left[\frac{i}{\hbar} 
\left(\frac{3}{\Lam} + 4\al \right) \left(k \pi_i + \frac{\pi_i^3}{27} \right) \right]
\notag \\
&&
\times
Ai\left[
\left(\frac{3}{\hbar^2 \Lam^2} \right)^{\frac{1}{3}} \left(3k - \Lam q_f \right)
\right] \, ,
\eea

\subsection{RBC at $t=0$}
\label{RBCt0}

We now focus on computing the expression in eq. (\ref{eq:Gamp})
for the case of Robin BC at the initial boundary. For the Robin BC the 
$S_{\rm tot}[q, N_c]$ is given in eq. (\ref{eq:Sact_frw_simp_RBC}). 
Using the expression mentioned in eq. (\ref{eq:Grbc_Gnbc}) one can 
relate the RBC path-integral with the NBC path-integral. This means we have 
\bea
\label{eq:GbarRBC_nbc_ms}
\int_{{\rm Bd}_i}^{{\rm Bd}_f} 
&&
{\cal D} q(t) \,\, 
\exp \left(\frac{i}{\hbar} S_{\rm tot}[q, N_c] \right)
= \exp\left[\frac{i}{\hbar} \left\{3k N_c  
+ 4 \al \left(k P_i + \frac{P_i^3}{27}\right) \right\}\right]
\notag \\
&& \times \left(\frac{2\pi\hbar}{i \bt}\right)^{1/2}
\int_{-\infty}^{\infty} \frac{{\rm d} \tilde{p}}{2\pi\hbar}
e^{i (P_i - \tilde{p})^2/2 \hbar \bt}
\bar{G}_{\rm NBC}[q_f, t=1; \tilde{p}, t=0] \, .
\eea
This when plugged in the full path-integral involving also 
the lapse $N_c$-integration gives
\bea
\label{eq:Gamp_rbc_ms_exp}
G_{\rm RBC}[{\rm Bd}_f, {\rm Bd}_i]&&
= \left(\frac{2\pi\hbar}{i \bt}\right)^{1/2} 
\exp\left[  
\frac{4 i \al}{\hbar}\left(k P_i + \frac{P_i^3}{27}\right) \right]
\int_{-\infty}^{\infty} \frac{{\rm d} \tilde{p}}{2\pi\hbar}
e^{i (P_i - \tilde{p})^2/2 \hbar \bt}
\notag \\
&&
\times
\int_{0^+}^{\infty} {\rm d} N_c 
\exp\left(\frac{3 k N_ci}{\hbar}\right)
\bar{G}_{\rm NBC}[q_f, t=1; \tilde{p}, t=0] \, .
\eea
Note that the $N_c$-integrand is not singular at $N_c=0$ 
allowing us to push the limit of integration all the way 
up to $-\infty$. This integral is however similar to the integral considered 
in the case of Neumann BC except for the absence of the Gauss-Bonnet
contribution. This means one can write 
\beq
\label{eq:Gbar_nbc_EH_part}
\int_{-\infty}^{\infty} {\rm d} N_c 
\exp\left(\frac{3 k N_ci}{\hbar}\right)
\bar{G}_{\rm NBC}[q_f, t=1; \tilde{p}, t=0]
= \biggl. \Psi_1(\tilde{p}) \biggr|_{\al =0} \times \Psi_2(q_f) \, ,
\eeq
where $\Psi_1$ and $\Psi_2$ are given in 
eq. (\ref{eq:psi1}) and (\ref{eq:psi2}) respectively. The 
$\Psi_2(q_f)$ can be computed as in case of NBC and is proportional to 
Airy-function. The function $\Psi_1(\tilde{p}) \biggr|_{\al =0}$ can also be 
computed by doing the following transformation 
\beq
\label{eq:pt_to_pb}
\tilde{p} \to \bar{p} - \frac{3 \Lam}{2\bt} \, ,
\eeq
which converts the original integral into an Airy-integral along with 
an exponential pre-factor.
Putting all the pieces together we have in the RBC case
\bea
\label{eq:Gamp_rbc_int_exact}
G_{\rm RBC}[{\rm Bd}_f, {\rm Bd}_i]
= &&
\sqrt{\frac{6\pi\hbar}{i \bt}}
\exp\left[  
\frac{4 i \al}{\hbar}\left(k P_i + \frac{P_i^3}{27}\right) \right]
\exp\biggl[
\frac{i\bigl(
-18 k \bt^2 + 2 P_i^2 \bt^2 
+ 6 P_i \bt \Lam + 3 \Lam^2
\bigr)}{4 \hbar \bt^3} 
\biggr] 
\notag \\
&&
\hspace{-15mm}
\times
\left(\frac{9}{\Lam\hbar}\right)^{\frac{1}{3}}
Ai\left[
\left(\frac{3}{\hbar^2 \Lam^2} \right)^{\frac{1}{3}} \left(3k - \Lam q_f \right)
\right] 
Ai \biggl[
\left(\frac{3\Lam}{\hbar^2}\right)^{\frac{1}{3}}
\left(\frac{3k}{\Lam}-\frac{P_{i}}{\bt}-\frac{3\Lam}{4\bt^2}\right)
\biggr] \, .
\eea
It is worth noting that for the RBC case also, the path-integral once again 
factorizes into two parts: one depends only on initial parameters ($P_i$ and $\beta$), and another only on final $q_f$. This factorization was known previously to hold at the semiclassical level. However, at the exact expression, it was first shown in \cite{Ailiga:2023wzl}.



\begin{figure}[h]
\centerline{
\vspace{0pt}
\centering
\includegraphics[width=10cm]{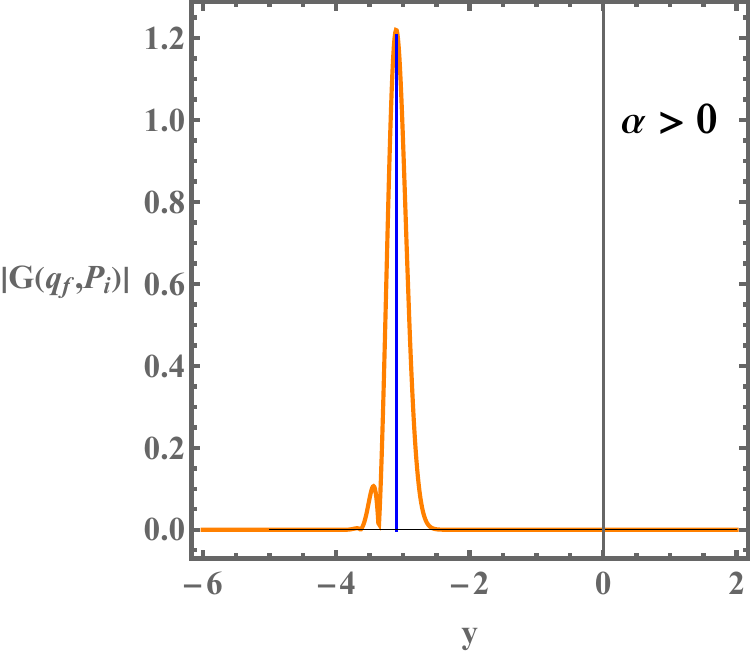}
}
\caption[]{
The plot of initial parameter dependent part of transition amplitude $G(q_f,P_i)$, given in Eq. (\ref{eq:Gamp_rbc_int_exact}). Here, we choose parameter values: $k=1$, $\Lam =3$, $q_f = 10$, $\hbar = 0.0001$, $\al = 1$ and $\beta = -1.5 i$. The $G(q_f,P_i)$ axis has been scaled appropriately for ease of visualization.
}
\label{fig:amp}
\end{figure}

For $\alpha > 0$, a constraint motivated from \cite{Cheung:2016wjt,Zwiebach:1985uq,Gross:1986mw,Metsaev:1987zx,Chakravarti:2022zeq}, it is noticed that the transition amplitude peaks around the initial parameter $ P = -3i$, 
which is  also the case, in absence of the Gauss-Bonnet for $\Lam > 0$.
This result is illustrated in Figure \ref{fig:amp}, where we plot the term dependent on initial parameters, $G_{\rm RBC}( q_f , P_{i})$, as a function of the Euclideanized variable \(P_i = i y\). This favours the Hartle-Hawking no boundary geometry as the most dominant geometry in the path integral.
In the subsequent section, we have shown it analytically. It is worth making the comparison with the NBC results in the limit 
$\bt\to0$, in which limit the RBC reduces to NBC. To see this, we use the following identity
\beq
\label{eq:beta_change}
\left(\frac{2\pi\hbar}{i \bt}\right)^{1/2} e^{i (P_i - \tilde{p})^2/2 \hbar \bt}
= \int_{-\infty}^\infty {\rm d} \xi \exp\left[
-\frac{i \bt}{2\hbar} \xi^2 + \frac{i}{\hbar} (P_i -  \tilde{p}) \xi
\right] \, .
\eeq
Here, one immediately notices that in the limit 
$\bt\to0$ one gets a $\de$-function $\de(P_i - \tilde{p})$. 
One can also obtain eq. \ref{eq:Gbd0bd1_full_nbc} from the 
eq. \ref{eq:beta_change} in the limit $\beta\rightarrow 0$, which involves 
asymptotic forms of the Airy-functions.

\section{$\hbar\to0$ limit}
\label{hbar0}

Having obtained the exact results for the transition amplitude, 
it is instructive to analyse the $\hbar\to0$ limit. This is crucial as one needs to compare it 
with the results obtained from the saddle-point approximation using Picard-Lefschetz 
methods (to be discussed later on). This limit also brings out the configurations which 
contribute dominantly to the path-integral and those which are suppressed. 
In our context, this translates into the initial values for the $\pi_i$ for NBC, 
$P_i$ and $\bt$ for the RBC. 

\subsection{Implications of having Gauss-Bonnet}
\label{GBcont}

\begin{figure}[h]
\centerline{
\vspace{0pt}
\centering
\includegraphics[width=9cm]{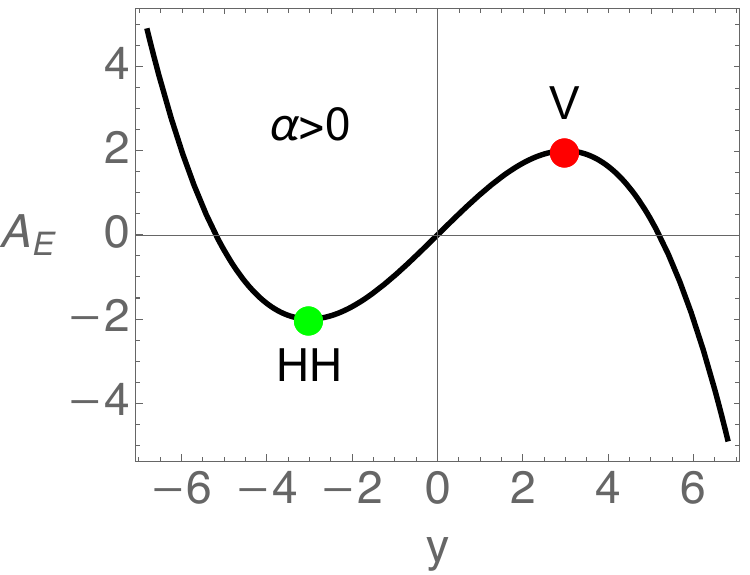}
}
\caption[]{
Plot of Euclidean action $A_E$ given in eq. (\ref{eq:Aeuc}) vs $y$. 
The two saddle points are depicted by green and red squares. 
For $\al>0$: green one corresponds to $P_i = -3 i \sqrt{k}$ (stable) 
while the red one corresponds to $P_i = 3 i \sqrt{k}$ (unstable).
\cite{Ailiga:2023wzl}
}
\label{fig:aept}
\end{figure}

Let's  focus on the exponential prefactor arising from the Gauss-Bonnet sector, which reads,
\beq
\label{eq:expGBfac}
\exp\left[  
\frac{4 i \al}{\hbar}\left(k P_i + \frac{P_i^3}{27}\right) \right]
\equiv e^{i A(P_i)/\hbar} \, ,
\eeq

In the limit $\hbar\to0$ the dominant contribution comes 
from geometries which extremize the exponent, which
corresponds to the 
\beq
\label{eq:exptA_pi}
A'(P_i) =0 \hspace{3mm} \Rightarrow \hspace{3mm}
P_i = \pm 3 i \sqrt{k} \, .
\eeq
In the limit $\bt\to0$ (NBC case) this corresponds to $\pi_i = \pm 3 i \sqrt{k}$! 
Two choices of signs correspond to two different choices for wick rotations.
Imaginary momentum depicts the quantum nature of the initial condition.
Let us consider taking $P_i = i y$, which 
is like considering a Euclideanised version action $A(P_i)$. 
Then we have the Euclidean action $A_E(y)$ which is given by
\beq
\label{eq:Aeuc}
A_E(y) = 4 \al \left(ky - \frac{y^3}{27} \right) \, ,
\hspace{5mm}
e^{i A(P_i)/\hbar}
\hspace{3mm} \Rightarrow \hspace{3mm}
e^{- A_E(y)/\hbar} \, .
\eeq
On extremization of Euclidean action $A_E(y)$ we observe that 
for $\al>0$
the point $y=-3\sqrt{k}$ corresponds to a stable saddle point leading to 
a positive exponent (aka Hartle-Hawking ) while 
$y=3\sqrt{k}$ corresponds to unstable saddle point leading to 
exponential with a negative argument (Vilenkin Tunneling). 
Coincidently, as we have shown in \ref{btchoice} and was also 
mentioned in \cite{DiTucci:2019bui, Narain:2022msz},
$P_i=-3i \sqrt{k}$ also corresponds to initial condition 
giving regular and well-behaved geometries within a certain regime of $\beta$. Since a universe with a higher positive value of the exponent is more probable, the choice $P_{i} = -3i\sqrt{k}$
is favored, as it maximizes the Gauss-Bonnet exponent.

Although, technically allowing \(\alpha < 0\) would result in the point \(P_i = 3i\sqrt{k}\) becoming a dominant geometry in the path integral, which would lead to a Vilenkin Tunneling proposal. However, the studies in string theory, such as those in \cite{Cheung:2016wjt,Zwiebach:1985uq, Gross:1986mw, Metsaev:1987zx}, as well as observational constraints from gravitational wave data and the area theorem, discussed in \cite{Chakravarti:2022zeq, Isi:2020tac}, limit the value of \(\alpha\) to positive values, restrict this possibility.
\subsection{Constraining $\bt$}
\label{btchoice}

Once $P_i$ gets fixed, the next thing that one has to do is to find the allowed possible 
values of parameter $\bt$ in the case of RBC transition amplitude. 
To find out the allowed range of $\bt$, one 
can analyse the nature of RBC transition amplitude in the $\hbar\to0$ limit. 
This can be achieved by focusing only on the part of the wave function 
which depends only on initial configuration.  

We start by plugging $P_i = -3i\sqrt{k}$ in eq. (\ref{eq:Gamp_rbc_int_exact})
and focusing first on the $\bt$-dependent Airy function. For $\bt$ negative 
imaginary ( as justified in section \ref{inter} ),
the Airy function has a positive argument for $\Lam>0$. This aids us 
in obtaining the $\hbar\to0$ limit by utilizing the asymptotic structure 
of the Airy-function
\beq
\label{eq:Phi_pi_hbar}
\left.
Ai \biggl[
\left(\frac{9\sqrt{\Lam}}{\hbar}\right)^{2/3}
\left(\sqrt{\frac{k}{\Lam}} + \frac{i \sqrt{\Lam}}{2\bt} \right)^2
\biggr] \right|_{\hbar \to 0}
\sim 
\exp\biggl[
- \frac{6\sqrt{\Lam}}{\hbar}
\left|\sqrt{\frac{k}{\Lam}} + \frac{i \sqrt{\Lam}}{2\bt} \right|^3
\biggr] 
= e^{
-B_1(\bt)/\hbar
}
\, .
\eeq
We next focus on the $\bt$-dependent 
exponential factor in eq. (\ref{eq:Gamp_rbc_int_exact}) and study its 
nature for $P_i = -3i\sqrt{k}$ in $\hbar\to0$ limit. For 
$P_i = -3i\sqrt{k}$ this exponential factor can be written 
as $e^{i B_2(\bt)/\hbar}$, where 
\beq
\label{eq:B2beta_func_def}
B_2(\bt)
=  \left(3\Lam^2 - 18 i \bt \Lam \sqrt{k} - 36 k \bt^2 \right)/(4 \bt^3) \, .
\eeq
Therefore, the overall exponential weighting for the RBC path integral 
in the semiclassical limit is expressed as:
\beq
\label{eq:Bbeta_exp_def}
\exp\biggl[
\frac{B(\bt)}{\hbar} 
\biggr]\equiv \exp\biggl[
\frac{B_1(\bt)+ B_2(\bt)}{\hbar} 
\biggr],
\eeq
From the 
structure, one could see that in limit $\hbar \rightarrow 0$ the dominant contribution comes from those
configurations for which $B^\prime(\bt_{\rm dom})=0$. This means we have
\beq
\label{eq:Bprime_bt}
B^\prime(\bt_{\rm dom}) = \frac{18i \Lam}{\bt_{\rm dom}^2}
\left(\sqrt{\frac{k}{\Lam}} + \frac{i \sqrt{\Lam}}{2\bt_{\rm dom}} \right)^2 = 0 \, .
\eeq
This gives the dominant configuration when $\bt$ lies around 
$\bt_{\rm dom} = -i \Lam /2\sqrt{k}$ (we don't 
consider the $\bt_{\rm dom}\to\infty$ case). At
this stage it is best to express 
\beq
\label{eq:beta_x_dep}
\bt = -\frac{i \Lam x}{2\sqrt{k}} 
= \bt_{\rm dom} x
\hspace{5mm} {\rm where} \hspace{5mm} x\geq 0 \, .
\eeq
This gives
\beq
\label{eq:xformB1B2}
B_1(\bt) = 6\frac{k^{3/2}}{\Lam} \biggl| 1 - \frac{1}{x} \biggr|^3 \, ,
\hspace{10mm}
B_2(\bt) = -\frac{6 i k^{3/2} \left(3x^2 -3x +1 \right)}{x^3 \Lam} \, .
\eeq
It immediately seen from here that there will be two cases in
this: $0\leq x \leq1$ and $x>1$. The dominant contributions 
comes from $x=1$ while $x=0$ refers to Neumann BC.

\subsubsection{$0\leq x \leq 1$}
\label{btlessBtdom}

In this range of $x$ simplification occurs and various terms in the 
exponentials combine to give
\beq
\label{eq:B1+B2_case1}
 \exp\biggl(
-\frac{ B_1(\bt)}{\hbar} 
\biggr)
\exp\left(\frac{i B_2(\bt)}{\hbar} \right)
= \exp\biggl[
\frac{6k^{3/2}}{\hbar \Lam}
\biggr] \, .
\eeq
When this is combined with the $\al$-dependent exponential pre-factor 
for $P_i=-3 i \sqrt{k}$ (which is given by $\exp(8k^{3/2} \al/\hbar)$)
one gets the following 
\beq
\label{eq:HHrbcFactor}
\exp\left(\frac{6 k^{3/2}}{\hbar\bar{\Lam}}\right)
\hspace{3mm} \text{where,} \,\, 
\bar{\Lam}= \Lam/\left(1+ 4\al \Lam/3\right) \, .
\eeq
This is the Hartle-Hawking state with positive weighting ($\Lambda>0$). 
The Gauss-Bonnet modification doesn't change the overall structure of 
the exponent. This further 
supports the findings of the paper \cite{Jonas:2020pos}, where 
it was noticed that Higher-derivative corrections don't prohibit the
no-boundary Universe as a solution. The analysis done in \cite{Jonas:2020pos} was
perturbative in nature. However, the topological nature of 
Gauss-Bonnet gravity helps us to go beyond 
the perturbative studies \cite{Ailiga:2023wzl}.  Although the contribution from Gauss-Bonnet appears as a multiplicative factor to the Einstein-Hilbert gravity, it is necessary to correctly reproduce the gravitation entropy of HDG theory \cite{Shu:2008yd}. It is known that the H-H exponent is equal to entropy/2 in Einstein's gravity. Our study also establishes this to be true for the higher-derivative gauss-bonnet gravity. Here, it is also seen that 
this exponential pre-factor is not only independent of the $\bt$-parameter
but also hint that a lower value of $\bar{\Lam}$ is more favourable. 

\subsubsection{$x>1$}
\label{btmoreBtdom}

Things are a bit different in the case when $x>1$. In this situation we have
\beq
\label{eq:B1+B2_case2}
 \exp\biggl(
-\frac{ B_1(\bt)}{\hbar} 
\biggr)
\exp\left(\frac{i B_2(\bt)}{\hbar} \right)
= \exp\biggl[
\frac{6k^{3/2}}{\hbar \Lam}
\left(
-1 + \frac{6}{x} - \frac{6}{x^2} + \frac{2}{x^3}
\right)
\biggr] \, .
\eeq
The total exponential factor also includes the factor coming from the 
Gauss-Bonnet sector $e^{8 k^{3/2} \al/\hbar}$. It is easy to see that 
in the limit $x\to\infty$, we get inverse Hartle-Hawking.
For any intermediate values $1<x<\infty$, the behaviour of exponential 
depends on the functional form of $x$. 
Suppose we call this function 
\beq
\label{eq:fx_beta_form}
f(x) = -1 + \frac{6}{x} - \frac{6}{x^2} + \frac{2}{x^3} \, .
\eeq
Then we notice that $f(1)=1$ and $f(\infty) = -1$. If we take 
derivative $f'(x) = - 6(x-1)^2/x^4 < 0$, then it is seen that it is 
always negative for all values of $x$ except $x=1$ and $x=\infty$.
It is a monotonically decreasing function of $x$ in the range 
$1$ and $-1$, but at some point $x_0$ the function becomes zero. 
This $x_0 = 2 + 2^{1/3} + 2^{2/3}$ is a crossover point. 
The function $f(x)>0$ for $1<x \leq x_0$, and $f(x)<0$ for 
$x>x_0$.

\subsection{Robin BC Interpretation}
\label{inter}

Before proceeding further, let us comment on a few things 
regarding why the Robin boundary condition at $t=0$ might be useful in 
the context of H-H no boundary proposal, which is a direct consequence 
of the above choice of $\beta$ (purely imaginary).
In this case, one can interpret 
the boundary terms as being coherent state {\it i.e.}, 
\begin{equation}\label{75}
   \exp\left[i\left(-\frac{\beta q_i^2}{2}
   +P_iq_i\right)\right] \xrightarrow[]{\text{for,\,\,} 
   \beta=-i|\beta|}\exp\left[-\frac{|\beta| q_i^2}{2}+iP_iq_i\right] \, .
\end{equation}
This is a coherent state with an imaginary 
momentum $P_i$ with the peak at $q_i=0$. One must be a little bit careful before making such an interpretation. Substituting naively $P_i=-ip$ with $p$ being real in eq. \ref{75}, one ends up with a real state with zero (mean) momentum with a peak at $q_i=p/|\beta|$. However, we should view the state in eq. \ref{75} as a complex generalization of the usual coherent state in the (analytically continued) phase space ($P_i,q_i$). With this in mind, eq.\ref{75} describes a state with shared uncertainty 
between the scale factor and the momentum 
of the universe at $t=0$. 
Taking this as an initial state \cite{DiTucci:2019bui, DiTucci:2019dji},
one can express the wave function as a gravitational path-integral 
in the following way
\begin{equation}
 \Psi[q_f,\beta,P_i] = \int dN 
 \mathcal{D}q \,dq_i e^{iS_{DD}[q_f,N,q]/\hbar} 
 \, \psi_0[P_i,\beta], \quad \psi_0[P_i,\beta] 
 \propto e^{i\left(- \frac{\beta q_i^2}{2}+P_iq_i\right)},
\end{equation}
where $S_{DD}$ is action with the Dirichlet BC
at the two endpoints. 
In the limit $\lvert \beta \rvert \to \infty$, 
$e^{-i \beta q_i^2 /2} \to \delta(q_i)$. This is a 
strict imposition of the Dirichlet BC $q_i=0$. The other
limit $\lvert \beta \rvert \to 0$ gives Neumann BC for which the state becomes a simple plane wave. 
Finite values of $\beta$ can be seen as a form of 'regulator',
embedding a regularized version of $\delta$-function 
in the path-integral. In a sense, RBC is a 
regularised DBC. 
This boundary condition is also compatible with the more fundamental quantum uncertainty 
principle in the sense 
knowing the initial size ($q_i$) arbitrarily accurate 
would render the initial 
momentum ($P_i$)  completely undetermined or vice versa. 
For example, the Neumann boundary condition at $t=0$ will allow 
the universe to start with an arbitrarily large size, which seems problematic to the no boundary philosophy. 
Moreover, quantum uncertainty doesn't allow
to determine the initial size (and/, or initial momentum) with arbitrary accuracy \cite{DiTucci:2019dji}. 
In a way, studying Robin BC is the 
most appropriate due to its compatibility 
with the uncertainty principle and the 
regularized way of introducing Dirichlet BC 
overcoming technical challenges associated with 
dealing delta function in path integrals.

\section{Picard-Lefschetz Methodology: mini review}
\label{PL}

The mini-superspace path-integral mentioned in eq. (\ref{eq:Gamp})
involves an integration over the real lapse $N_c$. Both in the NBC and 
RBC case one has to workout this integral either by directly computing it 
(as in NBC) or converting it into an integral over $\tilde{p}$ (as in RBC ). 
These integrals can also be worked out in saddle-point approximation 
and using Picard-Lefschetz methods. These methods helps us in analysing 
the behavior of the integrand 
in the complex plane \cite{Witten:2010cx,Witten:2010zr,Basar:2013eka,Tanizaki:2014xba}. 
This is achieved by determining the set of saddle points in the 
complex plane and finding the associated steepest ascent/descent paths. 
Then the Picard-Lefschetz methods systematically tell us the 
deformed integration contour along which the integrand is well-behaved 
and integral is convergent. To understand this in more detail, we consider a
generic oscillatory integral 
\beq
\label{eq:pathmock}
I = \int dz \, e^{i {\cal S}(z)/\hbar} \, ,
\eeq
where the exponent is a function of $z$ (real). Here $\hbar$ is a real parameter, 
and the action ${\cal S}(z)$ is a real-valued function. Generically, such integrals are not
absolutely convergent along the given contour. It may be either divergent or conditionally convergent. However, if it is conditionally 
convergent the Picard-Lefschetz (PL) methodology gives a systematic procedure 
for determining the deformed contour of integration along which the integral becomes absolutely convergent. In PL theory, 
one analytically continues both the integration parameter $z$ and the 
function ${\cal S}(z)$ into the complex plane. ${\cal S}(z)$ is 
then interpreted as an holomorphic function of $z$.
This implies that ${\cal S}$ satisfies the Cauchy-Riemann conditions
\begin{align}
\label{eq:CRfunc}
\frac{\de { S}}{\de \bar{z}} = 0 
\Rightarrow
\begin{cases}
\frac{\de {\rm Re} { S}}{\de x_1}
&= \frac{ \de {\rm Im} { S}}{\de x_2} \, , \\
\frac{\de {\rm Re} { S}}{\de x_2}
&= - \frac{ \de {\rm Im} { S}}{\de x_1} \, .
\end{cases}
\end{align}
%

\subsection{Flow equations}
\label{floweq}

If one writes the complex exponent as ${\cal I} = i {\cal S}/\hbar = h + iH$,
and expresses $z(\lambda) = x_1(\lambda) + i x_2(\lambda)$, then the downward flow can be written 
as
\beq
\label{eq:downFlowDef}
\frac{{\rm d} x_i}{{\rm d} \lam}
= - g_{ij} \frac{\pt h}{\pt x_j} \, ,
\eeq
where the metric $g_{ij}$ appearing above is a Riemannian 
metric defined on the complex plane, $\lam$ is the flow 
parameter, and the $(-)$ sign represents the downward flow.
The solution to these equations gives the steepest descent 
trajectories associated with the corresponding saddle point (also known as critical points). 
These are also known as \textit{thimbles} 
(denoted by ${\cal J}_\sg$, $\sigma$ subscript symbolizes saddles). If, in the above set of equations, 
we have a $(+)$ sign instead of $(-)$, then we have the 
equations for the steepest ascent flow lines (denoted by ${\cal K}_\sg$). The flow parameter ($\lambda$) takes the values in the range of ($-\infty$,$+\infty$). For a downward flow emanating from the saddle, $\lambda$ takes $-\infty$ at the saddles and increases as we move along the flow line away from the saddles. Along the steepest ascent line, $\lambda$ increases as we move towards the saddle point and takes $+\infty$ at the saddles.
The subscript $\sg$ refers to the saddle point to which these sets of 
trajectories are attached. From this definition, it is quickly seen that 
the real part $h$ (also called Morse function) decreases 
monotonically along the steepest descent lines while increasing 
monotonically along the steepest ascent ones. To confirm this 
one can consider 
\beq
\label{eq:flowMonoDec_h}
\frac{{\rm d} h}{{\rm d} \lam}
= g_{ij} \frac{{\rm d} x^i}{{\rm d} \lam} \frac{\pt h}{\pt x_j} 
= - \left(\frac{{\rm d} x_i}{{\rm d}\lam}\frac{{\rm d} x^i}{{\rm d}\lam}\right)
\leq 0 \, .
\eeq
This holds true for any generic Riemannian metric (matrices which have a positive inner product). However, for simplicity 
and understanding the nature of these equations, it is useful to adopt a 
metric (Kahler metric). In our case, we'll opt 
for the flat Cartesian metric, $ds^2 = \lvert dx \rvert^2$. 
With complex coordinates $(z,\bar{z} = (x+iy,x-iy))$, 
the metric components are $g_{z,z}=g_{\bar{z},\bar{z}}=0$
and $g_{z,\bar{z}}= g_{\bar{z},z}=1/2$. This simplifies flow equations as:
\beq
\label{eq:simpflow}
\frac{{\rm d}z}{{\rm d} \lam} = \pm \frac{\pt \bar{\cal I}}{\pt \bar{z}} \, ,
\hspace{5mm}
\frac{{\rm d}\bar{z}}{{\rm d} \lam} = \pm \frac{\pt {\cal I}}{\pt z} \, .
\eeq
From this set of equations, one notices that the imaginary 
part of ${\rm Im} {\cal I}=H$ is constant along the flow lines. 
\beq
\label{eq:consHflow}
\frac{{\rm d}H}{{\rm d} \lam} 
= \frac{1}{2i} \frac{{\rm d} ({\cal I} - {\cal \bar{I}})}{{\rm d} \lam} 
= \frac{1}{2i} \left(
\frac{\pt {\cal I}}{\pt z} \frac{{\rm d} z}{{\rm d} \lam} 
- \frac{\pt \bar{\cal I}}{\pt \bar{z}}\frac{{\rm d}\bar{z}}{{\rm d} \lam}
\right) = 0 \, .
\eeq
This feature can be exploited to find the flow trajectories, which otherwise 
maybe hard to find out as one has to deal with the 
differential equations. However, in order to find the nature of 
trajectory (whether steepest ascent/descent), one just needs to 
analyse the behaviour of ${\rm d} h(\lam)/{\rm d} \lam$. 
Note that for real ${\cal S}(z)$, the morse function $h$ will vanish
on the real line. This means that $h$ along the steepest ascent trajectories 
which intersect the real line will be more negative when traced backwards. 
Hence the saddle points whose associated steepest ascent line intersects the 
real line (original integration contour) will have negative $h$. Interestingly, the integration done along the 
thimbles of this saddle will be convergent by definition.  These saddles 
are called \textit{relevant}. In a generic situation, when $S(z)$ is not necessarily real to start with, only those saddles will be called \textit{relevant} from which the associated steepest ascent lines intersect the original integration contour. These saddles contribute to the gravitational path integral and hence are called \textit{relevant}.

\subsection{Choice of contour}
\label{choice}

Once the saddle points and associated flow lines have been computed, 
one needs to work out a way to deform the original integration contour and 
make the integral absolutely convergent 
(for more detail see \cite{Witten:2010zr,Basar:2013eka,Feldbrugge:2017kzv}).

In the complex plane, the behaviour of $h$ and $H$ will aid us in determining 
the correct deformed contour of integration. If one denotes a particular saddle 
as $N_\sg$ then the region is called `allowed' if for each point $N$ 
in the region we have $h(N) < h(N_\sg)$. This `allowed' region is denoted by $J_\sg$.
A region is called `forbidden' if, for each point $N$ in the region, we have 
$h(N)>h(N_\sg)$. The `forbidden' region is denoted by $K_\sg$. 
Intuitively, this can be thought of as a landscape ridded with valleys (allowed region),
hills (forbidden region) and plains (intermediate region). 

It is also important to note that $h(\lam) \to -\infty$ along the steepest 
descent lines ($\mathcal{J}_\sigma$) when $\lam \to + \infty$ (i.e., moving away from the saddles), while $h(\lam) \to \infty$
along the steepest ascent ($\mathcal{K}_\sigma$) lines when $\lam \to - \infty$ (i.e., moving away from the saddles). This implies that the integrand is dying out as we move away from the saddles along the steepest descent thimbles, and the integral along the whole thimble is dominated by the saddle point contribution only. This is what we need to validate the saddle point approximation. However, along the steepest ascent lines, the integrand increases exponentially as we move away from the saddle points, and this invalidates the saddle point approximation.
These lines ($\mathcal{J}_\sigma$ and $\mathcal{K}_\sigma$) intersect only at the saddle point and where 
they satisfy\footnote{ We have already removed the degeneracies related to the stokes ray which connects two saddle points. This happens when the steepest descent curve from a saddle point meets the steepest ascent curve from another saddle point as a result of symmetry. This can be removed by adding a small perturbation (off-shell fluctuation) to the exponent.}
\beq
\label{eq:JKorient}
{\rm Int} \left({\cal J}_\sg, {\cal K}_{\sg^\prime} \right) = \de_{\sg \sg^\prime} \, ,
\eeq
where ``$\rm Int$" counts the intersection between two curves. Now let us deform the original integration contour ($\mathbb{D}$) to the deformed contour ($\mathcal{C}$) and write  $\mathcal{C}=\sum_\sigma n_\sigma \mathcal{J}_\sigma$,
for some integers $n_\sg$ which will take 
value $0$ or $\pm1$ depending on the orientation. This gives
$n_\sg = {\rm Int} ({\cal C}, {\cal K}_\sg) = 
{\rm Int} (\mathbb{D}, {\cal K}_\sg)$. Being topological 
$n_\sg$ doesn't change if one deforms the contours. 
It also implies that the necessary and sufficient condition for a saddle to be 
\textit{relevant} if the steepest ascent emanating from the saddle intersects 
the original integration contour for which only $n_\sigma$ is non-zero. The corresponding thimble 
${\cal J}_\sg$ becomes the deformed integration contour. It is also crucial that the deformed 
integration contour shouldn't lie in the forbidden region near the endpoints of the original integration, which we always keep fixed during the deformation. This ensures 
that we can safely slide the original integration contour 
into the deformed integration contour by cauchy's theorem. 

It is often the case that more than one saddle is relevant, and hence there will be multiple thimbles that one would need to 
integrate over. This will imply that the original integral has been 
converted into integrals over all the relevant thimbles
\beq
\label{eq:sumOthim}
I = \int_{\cal C} {\rm d} z e^{i \mathcal{S}(z)/\hbar}
= \sum_\sg n_\sg \int_{{\cal J}_\sg} {\rm d}z
e^{i \mathcal{S}(z)/\hbar} \, .
\eeq
This is the feature of the Picard-Lefschetz methodology. 
The integration done along each of the thimble is absolutely convergent
as can be seen from
\beq
\label{eq:absconvg}
\biggl| \int_{{\cal J}_\sg} {\rm d} z
e^{i\mathcal{S}(z)/\hbar} \biggr| \leq
\int_{{\cal J}_\sg} \lvert {\rm d} z \rvert
\lvert e^{i \mathcal{S}(z)/\hbar} \rvert
= \int_{{\cal J}_\sg} \lvert {\rm d} z \rvert e^h(z) < \infty \, .
\eeq
Denoting the `distance' along the contour path as $l= \int \lvert {\rm d} z \rvert$.
Then for the above integral to converge it is necessary that 
$e^h \sim 1/l$ as $l\to \infty$. Consequently, the original integration can 
be analytically deformed into a sum of absolutely convergent integrals 
along various thimbles passing through the relevant saddle points. 
By expanding in $\hbar$, the leading-order expression is as follows:
\beq
\label{eq:LDordI}
I = \int_{\cal C} {\rm d} z e^{i \mathcal{S}(z)/\hbar}
= \sum_\sg n_\sg e^{i H(z_\sg)} \int_{{\cal J}_\sg} {\rm d}z e^{h} 
\approx \sum_\sg n_\sg e^{i \mathcal{S}(z_\sg)/\hbar} \left[{\cal A}_\sg 
+ {\cal O}(\hbar) \right] \, .
\eeq
Here, ${\cal A}_\sigma$ represents the contribution 
(fluctuation) obtained after performing a Gaussian integration 
around the saddle point $z_\sigma$.

\section{Saddle point approximation}
\label{spa} 

Having detoured a bit to learn about the Picard-Lehschetz, we return back to 
analyzing the mini-superspace path-integral. To proceed further and gain deeper 
understanding of the results obtained so far it is best to do saddle-point studies 
and compare it with the exact results and those obtained under the $\hbar\to0$ limit.
The $\hbar\to0$ study has shown us that the dominant contribution in the 
path-integral comes from the $P_i= \pm3 i \sqrt{k}$, and only 
$P_i = -3 i \sqrt{k}$ correspond to stable configuration. This is also the 
value of $P_i$, which leads to stable behaviour of perturbations 
in the Hartle-Hawking no-boundary proposal of the Universe
\cite{DiTucci:2019bui,Narain:2021bff, Lehners:2021jmv, 
DiTucci:2020weq, DiTucci:2019dji, Narain:2022msz}. 
It is important to highlight that in a sense the Gauss-Bonnet 
favours the Hartle-Hawking no-boundary Universe. 

We will now do the saddle point analysis of the of the 
mini-superspace path-integral for the RBC case 
(we will skip the NBC analysis which is presented in \cite{Narain:2022msz}).
However, we will not apply the Picard-Lefschetz methods directly 
on the $N_c$-integral, instead as the RBC path-integral can be written 
as an NBC path-integral using eq. (\ref{eq:Gamp_rbc_ms_exp})
and (\ref{eq:Gbar_nbc_EH_part}), one can then study the 
integral over the $\tilde{p}$ and $N_c$ using PL separately. 

In the integral over $N_c$ mentioned in eq. (\ref{eq:Gbar_nbc_EH_part}) if 
one makes a shift as in eq. (\ref{eq:NcTONb}) then it separates 
in two parts: one dependent on $\tilde{p}$ and other an integral 
over $\bar{N}$ (not dependent on $\tilde{p}$). 
The $\bar{N}$-integral depends on $q_f$ is written as 
$\Psi_2(q_f)$, and its saddle analysis has been done in 
\cite{Narain:2022msz} (won't be included here). It was noticed that 
for all values of $q_f$ there are two saddle points whose nature changes 
for $q_f<3k/\Lam$, $q_f=3k/\Lam$ and $q_f>3k/\Lam$. 
In the first case both saddles lie on imaginary axis however only one
of them is relevant. In the case when $q_f>3k/\Lam$, both 
saddle are complex in nature and both are relevant. 
The case of $q_f=3k/\Lam$ is a degenerate case. 
For $q_f<3k/\Lam$, Universe is Euclidean as can be seen from the 
exponential growth of the transition amplitude referring to 
imaginary time, while in the $q_f>3k/\Lam$ the transition amplitude is 
oscillatory in nature indicating the emergence of real time and a Lorentzian Universe. 

We then move to analysing the behaviour of the $\tilde{p}$-integral. 
This integral is given by
\beq
\label{eq:Bint_def_Bact}
\int_{-\infty}^{\infty} \frac{{\rm d} \tilde{p}}{2\pi\hbar}
e^{i (P_i - \tilde{p})^2/2 \hbar \bt} \biggl. \Psi_1(\tilde{p}) \biggr|_{\al =0}
= \int_{-\infty}^{\infty} \frac{{\rm d} \tilde{p}}{2\pi\hbar}
e^{i B(\tilde{p})/\hbar}
=  \int_{-\infty}^{\infty} \frac{{\rm d} \tilde{p}}{2\pi\hbar}
e^{\{h(\tilde{p}) + i H(\tilde{p}) \}/\hbar} \, ,
\eeq
where
\beq
\label{eq:argExp_Pt_phi}
B(\tilde{p}) = \frac{(P_i - \tilde{p})^2}{2\bt} + \frac{3}{\Lam} 
\left(
k \tilde{p} + \frac{\tilde{p}^3}{27}
\right) \, .
\eeq
and $h(\tilde{p})$ is the corresponding Morse-function while the 
$H(\tilde{p})$ corresponds to real part of the $B$-function. Once again 
one can recast this into an Airy-integral by a shift of variable
as discussed previously. Here we study this using 
Picard-Lefschetz methods (see \cite{Feldbrugge:2017kzv, Narain:2021bff, 
Witten:2010cx, Witten:2010zr, Basar:2013eka, Tanizaki:2014xba} 
for review on Picard-Lefschtez and analytic continuation). 

The saddle-points of $\tilde{p}$ can be obtained by by 
setting ${\rm d} B(\tilde{p})/{\rm d} \tilde{p}=0$. These are given by
\beq
\label{eq:pt_sads_12_HH}
\tilde{p}_1 = - 3 i \sqrt{k} \, ,
\hspace{5mm}
\tilde{p}_2 = \frac{3i \sqrt{k} (x-2)}{x} \, .
\eeq
This shows that while saddle point $\tilde{p}_1$ remains fixed at 
the same position for all $x$, the saddle point $\tilde{p}_2$ moves 
as $x$ changes. It becomes zero for $x=2$. At the saddle point 
$\tilde{p}_1$, we have a vanishing initial on-shell geometry, i.e., $\bar{q}_i=0$, 
while at the other saddle point $\tilde{p}_2$, we have a non-vanishing 
initial geometry $\bar{q}_i = 12 k (x-1)/\Lam x^2$. 

One can expand the function $B(\tilde{p})$ around the saddle point $\tilde{p}_\sg$
where $\sg = \{1,2\}$. This gives
\beq
\label{eq:Bexp_ptsig}
B(\tilde{p}) = B(\tilde{p}_\sg)
+ \biggl. \frac{{\rm d} B(\tilde{p})}{{\rm d} \tilde{p}} \biggr|_{\tilde{p}_\sg} \de \tilde{p}
+ \frac{1}{2} \biggl. \frac{{\rm d}^2 B(\tilde{p})}{{\rm d} \tilde{p}^2} \biggr|_{\tilde{p}_\sg} (\de \tilde{p})^2
+  \frac{1}{6} \biggl. \frac{{\rm d}^3 B(\tilde{p})}{{\rm d} \tilde{p}^3} \biggr|_{\tilde{p}_\sg} (\de \tilde{p})^3 \, ,
\eeq
where $\de \tilde{p} = \tilde{p} - \tilde{p}_\sg$.
The second variation at the saddle-point can be written 
as $B''(\tilde{p}_\sg)= r_\sg e^{i \rho_\sg}$,
where $r_\sg$ and $\rho_\sg$ depends on boundary conditions. 
Near the saddle point, the change in $i H$ will go like 
\beq
\label{eq:changeH}
\de (iH) \propto i 
\left(B''(\tilde{p}_\sg) \right) \left(\de \tilde{p} \right)^2
\sim v_\sg^2 e^{i\left(\pi/2 + 2\ta_\sg + \rho_\sg \right)} \, ,
\eeq
where we write $\de \tilde{p} = v_\sg e^{i\ta_\sg}$ and $\ta_\sg$ 
is the direction of flow lines at $\tilde{p}_\sg$ and given by
\beq
\label{eq:flowang}
\ta_\sg = \frac{(2k-1)\pi}{4} - \frac{\rho_\sg}{2} \, ,
\eeq
where $k \in \mathbb{Z}$. 

For the steepest descent and ascent flow lines, their 
corresponding $\ta_\sg^{\rm des/aes}$ is such that the phase for 
$\de H$ correspond to $e^{i\left(\pi/2 + 2\ta_\sg + \rho_\sg \right)} = \mp1$. This implies
\beq
\label{eq:TaDesAes}
\ta_\sg^{\rm des} = k \pi + \frac{\pi}{4} - \frac{\rho_\sg}{2} \, ,
\hspace{5mm}
\ta_\sg^{\rm aes} = k \pi - \frac{\pi}{4} - \frac{\rho_\sg}{2} \, .
\eeq
These angles can be computed numerically in our case.  

Under the saddle point approximation, the contour integral given in 
eq. (\ref{eq:Bint_def_Bact}) can be computed using Picard-Lefschetz methods
(see \cite{Narain:2021bff} for details of PL methodology).
This gives
\bea
\label{eq:Bint_sadPL_apr}
\int_{-\infty}^{\infty} \frac{{\rm d} \tilde{p}}{2\pi\hbar}
e^{i B(\tilde{p})/\hbar}
&&
= \sum_\sg \frac{n_\sg}{2\pi\hbar} e^{i B(\tilde{p}_\sg)/\hbar}
\int_{{\cal J}_\sg} {\rm d} \de\tilde{p}
\exp\left[i \frac{B''(\tilde{p}_\sg)}{2\hbar} (\de \tilde{p})^2 \right] \, ,
\notag \\
&&
= \sum_\sg n_\sg \sqrt{\frac{1}{2\pi\hbar \lvert B''(\tilde{p}_\sg) \rvert}}
e^{i\ta_\sg} e^{i B(\tilde{p}_\sg)/\hbar} \, .
\eea
where ${\cal J}_\sg$ refers to steepest descent line/thimble
and $n_\sg$ is the intersection number which will take
values ($0, \pm1$) accounting for the orientation of contour over each thimble.
In the following 
we will be use this expression to compute the contour integral 
for various values of $x$.

\subsection{$0 < x < 1$}
\label{BsadsXless1}

For $0< x<1$, both the saddle points $\tilde{p}_1$ and $\tilde{p}_2$ lie on the negative
imaginary axis with $\lvert \tilde{p}_2 \rvert< \lvert \tilde{p}_1\rvert$.

\begin{figure}[h]
\centerline{
\vspace{0pt}
\centering
\includegraphics[width=3in]{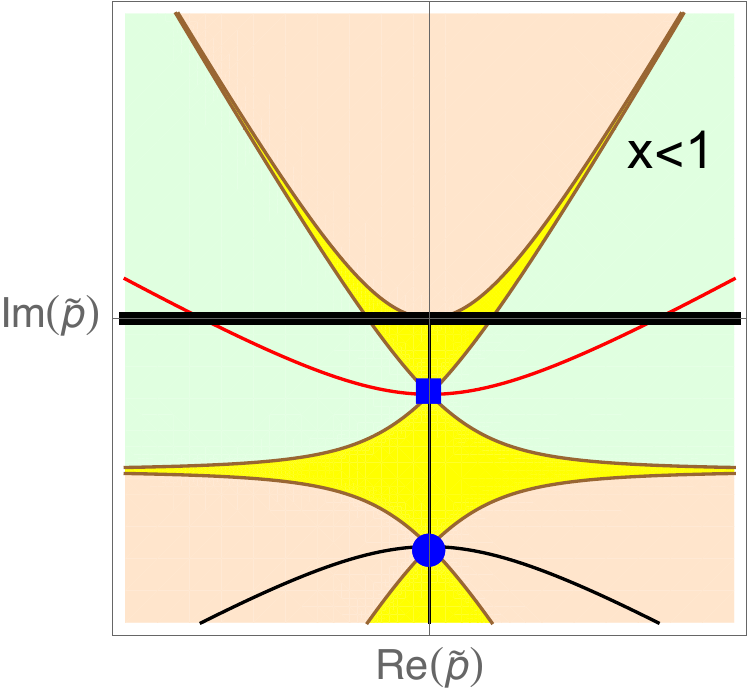}
}
\caption[]{
This figure depicts the Picard-Lefschetz analysis of the 
contour integral given in eq. (\ref{eq:Bint_def_Bact}).
The red lines correspond to the steepest descent lines (thimbles ${\cal J}_\sg$), while 
the thin black lines are the steepest ascent lines and are denoted by ${\cal K}_\sg$. 
Here, we choose parameter values: $k=1$, $\Lam=3$, and $x=1/2$. 
For this, the saddle point $\tilde{p}_1$ corresponds to the blue square, and it is relevant, while 
saddle point $\tilde{p}_2$ corresponds to the blue circle.
The thick black line shows the original integration contour 
$(-\infty, +\infty)$. The light-green region is the allowed region 
with $h<h(\tilde{p}_\sg)$ for all values of $\sg$. The light-orange region 
(forbidden region) has $h>h(\tilde{p}_\sg)$ for all $\sg$. 
The intermediate region is depicted in yellow \cite{Ailiga:2023wzl}.
}
\label{fig:Bactsad_xL1}
\end{figure}

In figure \ref{fig:Bactsad_xL1}, we plot the various 
flow-line, saddle points, and forbidden/allowed regions. 
From the graph, we notice that only the steepest ascent line from 
$\tilde{p}_1$ intersects the original integration contour (real line), 
thereby making it relevant. This contour picks the correct Airy-function 
to be $Ai(.)$. At this saddle point, the initial geometry turns out to be 
vanishing, thereby satisfying both the ``compactness'' and ``regularity" 
criterion of Hartle-Hawking's no-boundary proposal. 
The Picard-Lefschetz theory then gives the following  
in the saddle point approximation as
\beq
\label{eq:transamp_PL_xL1}
\int_{-\infty}^{\infty} \frac{{\rm d} \tilde{p}}{2\pi\hbar}
e^{i (P_i - \tilde{p})^2/2 \hbar \bt} \biggl. \Psi_1(\tilde{p}) \biggr|_{\al =0}
\approx \sqrt{\frac{ \Lam x k^{-1/2}}{4\hbar\pi (1-x)} } \,\,
e^{\frac{6 k^{3/2}}{\hbar \Lam}} \, .
\eeq
Hence, we get the same exponential factor as in the Hartle-Hawking
no-boundary Universe. 
Gauss-Bonnet being topological in 4D won’t alter this as long as perturbations are small.

\subsection{$x>1$}
\label{BsadsXmore1}

For $x>1$, only saddle 
point $\tilde{p}_1$ remains fixed on the negative imaginary axis for 
all values of $x$, while the saddle point $\tilde{p}_2$ moves from
negative imaginary axis to positive imaginary axis. 

\begin{figure}[h]
\centerline{
\vspace{0pt}
\centering
\includegraphics[width=2.7in]{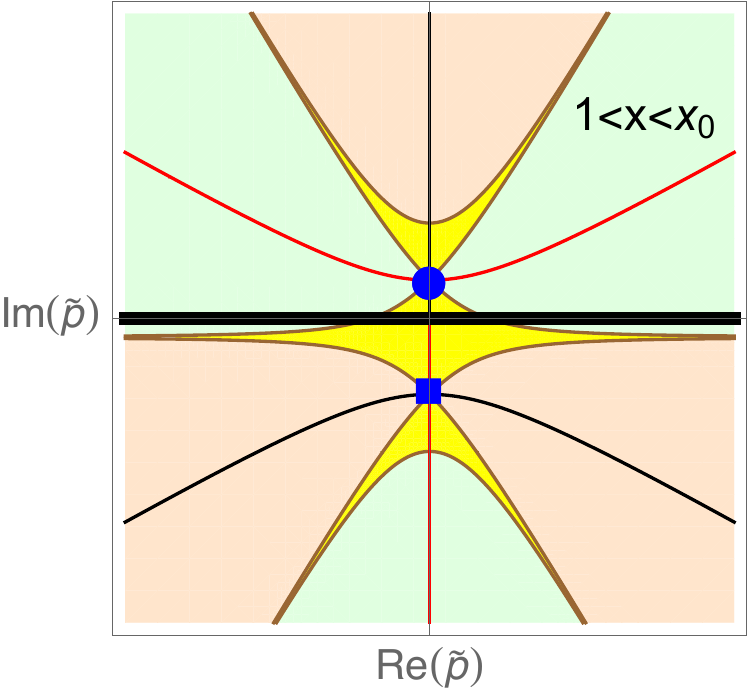}
\hspace{10mm}
\includegraphics[width=2.7in]{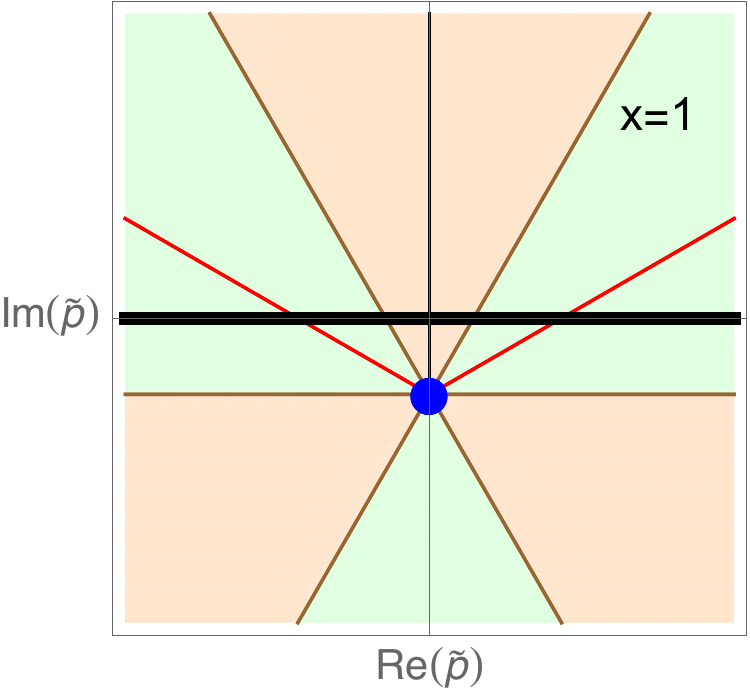}
}
\caption[]{ Here we choose parameter values: $k=1$, $\Lam=3$, and $x=4$ (left) and $x=1$ (right) respectively. 
For the left figure, the saddle point $\tilde{p}_2$ corresponds to blue- circle, while 
saddle point $\tilde{p}_1$ corresponds to blue-square. Only the 
saddle point $\tilde{p}_2$ (blue square) is relevant. 
The right figure corresponds to the degenerate case when 
two saddle points merge, and they are both relevant. 
}
\label{fig:Bactsad_xM1}
\end{figure}

We see that only the steepest ascent 
curves emanating from $\tilde{p}_2$ intersect the 
original integration contour, implying that it is the 
only relevant saddle point. At this saddle, the initial geometry of the 
universe is non-vanishing, given by $\bar{q}_i= 12 k (x-1)/{\Lam x^2}$. 
The descent thimble passing through this saddle will be the deformed 
contour of integration, which again picks $Ai(.)$ in eq. (\ref{eq:Bint_def_Bact}).
The Picard-Lefschetz analysis then gives the following  
in the saddle point approximation as
\beq
\label{eq:transamp_PL_xM1}
\int_{-\infty}^{\infty} \frac{{\rm d} \tilde{p}}{2\pi\hbar}
e^{i (P_i - \tilde{p})^2/2 \hbar \bt} \biggl. \Psi_1(\tilde{p}) \biggr|_{\al =0}
\approx 
\sqrt{\frac{ \Lam x k^{-1/2}}{4\pi\hbar (x-1)} } \,\,
e^{\frac{6 k^{3/2}}{\hbar \Lam}\left(
\frac{2}{x^3} - \frac{6}{x^2} + \frac{6}{x} -1
\right)} \, .
\eeq
%

\subsection{$x=1$}
\label{BsadsX=1}

This is the degenerate case where both the
saddle points merge : $\tilde{p}_1 = \tilde{p}_2 = -3 i \sqrt{k}$, 
with vanishing initial geometry at the relevant saddle. 
In this case the Morse-function becomes $h(\tilde{p}_1) = h(\tilde{p}_2)
= 6k^{3/2}/\Lam$. 
Saddle-point approximation breaks down 
as the second derivative vanishes, $B''(\tilde{p}_\sg)=0 $. 
One needs to go third order in the series expansion in 
eq. (\ref{eq:Bexp_ptsig}).
This situation is depicted in figure \ref{fig:Bactsad_xM1}.
In this case, the integration gives the following 
\beq
\label{eq:transamp_PL_xDE1}
\int_{-\infty}^{\infty} \frac{{\rm d} \tilde{p}}{2\pi\hbar}
e^{i (P_i - \tilde{p})^2/2 \hbar \bt} \biggl. \Psi_1(\tilde{p}) \biggr|_{\al =0}
\approx 
\frac{\sqrt{3}}{2\pi\hbar} e^{iB(\tilde{p}_\sg)/\hbar} \int_0^\infty {\rm d} v_\sg
e^{-v_\sg^3/9\Lam \hbar}
= \frac{(\Lam /3\hbar^2)^{1/3}}{\Gamma(2/3)}
e^{\frac{6 k^{3/2}}{\hbar \Lam}} \, .
\eeq
%

\section{Connection to WDW equation}
\label{WDW1}

So far, we have been focusing on the path integral quantization. 
Employing the path integral approach is both technically and conceptually 
advantageous, especially when the initial condition is specified on the geometry, 
such as in this case, on the metric variable, rather than directly on the 
wave-function (as in the case of the DeWitt condition).  However, one can ask how
these results are related to the solution of the Wheeler-DeWitt equation in the canonical approach.

For the mini-superspace model eq. (\ref{eq:frwmet_changed}), the Lagrangian is given by:
\beq\label{eq:Lag}
L = - \frac{3}{4 N}\dot{q}^2 + 3 N - N \Lam q \, ,
\eeq
where we have set $k=1$ and done integration by parts in eq. (\ref{eq:Sact_frw_simp})
to obtain the above Lagrangian.
The canonical momentum $p$ conjugate to $q$ is:
\beq\label{eq:mom}
p = \frac{\partial L}{\partial q}=- \frac{3}{2N}\dot{q}.
\eeq
Consequently, the Hamiltonian takes the following form
\beq
\label{EH_Ham}
{H}_{\rm {EH}} = \dot{q}p - L = N \left[- \frac{p^2}{3}+(\Lam q - 3k) \right]= N \hat{H}_{EH} \, .
\eeq
The Gauss-Bonnet Hamiltonian in 4 spacetime dimensions 
is zero \cite{Liu:2008zf} i.e., ${H}_{GB}=0$. This is because the Gauss-Bonnet term is topological in four dimensions 
and should not affect the bulk dynamics.

The Hamiltonian describing the Einstein-Gauss-Bonnet model is given by 
\beq 
\label{Ham}
\hat{H}_{EGB} = \hat{H}_{\rm {EH}} + \hat{H}_{\rm {GB}} = \hat{H}_{\rm {EH}} \, .
\eeq
The Wheeler-DeWitt equation corresponds to the operator version of classical constraint $H_{EGB}=0$:
\beq
\label{WDW}
\hat{H}_{EGB} \Psi = 0 \, ,
\eeq
where $p$, $q$ appearing in eq. (\ref{EH_Ham}) have been promoted to operators with the commutation relation $[\hat{q}, \hat{p}]=i\hbar$. 
Eq. \ref{WDW} implies that physical states are annihilated by the  Hamiltonian.

In the following, we will employ this equation at $t=0$, 
where we impose the robin boundary condition $p_i+\beta q_i=P_i,q_i=Q_i$. 
Consequently, the Hamiltonian constraint takes the form
\beq
\left[-\frac{1}{3}\left(P_i-\beta Q_i \right)^2+\left(\Lambda Q_i-3\right)\right]=0.
\eeq
In order to derive the Wheeler-DeWitt (WDW) equation, 
we promote the variables to operators. It will be convenient to transition 
into momentum space by substituting 
$\hat{Q}_i=i\hbar\frac{\partial}{\partial P_i}$ and $\hat{P}_i=P_i$. 
Then, the WDW equation reads
\beq
\label{wdw}
\frac{1}{3}\left(P_i^2\Psi_0+i\hbar\beta P_i\frac{\partial \Psi_0}{\partial P_i}
+i\hbar\beta \frac{\partial}{\partial P_i}(P_i\Psi_0)
-\beta^2\hbar^2\frac{\partial^2\Psi_0}{\partial P_i^2}\right)
+\left(i\Lambda\hbar\frac{\partial\Psi_0}{\partial P_i}+3\Psi_0\right)=0
\eeq
where $\Psi_0=\Psi_0(P_i)$. Note that in eq. \ref{wdw}, 
we flip the sign $\frac{\partial}{\partial P_i}\rightarrow-\frac{\partial}{\partial P_i}$ 
because we are considering this at the initial boundary. 
It is a second-order differential equation. The most general solution is 
given by\footnote{The Airy functions $Ai$ and $Bi$ are defined as two 
linearly independent solutions. Rotating the arguments by $2\pi/3$ results different linear combinations of Airy functions, given by: 
$Ai(ze^{\pm i \frac{2\pi}{3}}=\frac{1}{2}e^{\pm i\frac{\pi}{3}}\left[Ai(z)\mp i Bi(z)\right]$}:
\beq
\begin{split}
\Psi_0(P_i)= \exp\left[i\left(\frac{P_i^2}{2\hbar\beta}
+\frac{3P_i\Lambda}{2\hbar \beta^2}\right)\right]\times\left[c_1 Ai \biggl[
\left(\frac{-3\Lam}{\hbar^2}\right)^{\frac{1}{3}}
\left(-\frac{3}{\Lam}+\frac{P_{i}}{\bt}+\frac{3\Lam}{4\bt^2}\right)
\biggr]\right.\\
+c_2 \left. Bi\biggl[
\left(\frac{-3\Lam}{\hbar^2}\right)^{\frac{1}{3}}
\left(-\frac{3}{\Lam}+\frac{P_{i}}{\bt}+\frac{3\Lam}{4\bt^2}\right)
\biggr]\right]
\end{split}
\eeq
where, $c_1$ and $c_2$ are arbitrary functions of $\beta$ 
and $\Lambda$. To match with the $t=0$ 
part of the path-integral expression in eq. \ref{eq:Gamp_rbc_int_exact}, 
we choose the following,
\beq
\begin{split}
c_1=-ic_2=\frac{1}{2}\sqrt{\frac{3\pi \hbar}{2i\beta}}
e^{-i\pi/3}\exp\left[i\left(-\frac{9}{2\hbar \beta}+\frac{3\Lambda^2}{4\hbar \beta^3}\right)\right]
\end{split}
\eeq
This fixes the constants $c_1$ and $c_2$ uniquely. 
At this stage, it's worth noting a basic difference between 
$\beta=0$ (Neumann condition) and $\beta\neq 0$ cases. 
For $\beta=0$, the Wheeler-DeWitt (WDW) equation is a first-order 
differential equation and hence has only one solution. In this scenario, 
the WDW equation provides the correct wave-function without any 
reference to the path-integral. The reason behind this is that computing 
the $t=0$ part from the path-integral doesn't require any specific choice of 
contour. The lesson learned is that the specific choice of contour directly 
relates to the arbitrary constants generated in the WDW equation. 
At $t=1$, it's convenient to work in position space by substituting 
$p_f=-i\hbar \frac{\partial}{\partial q_f}$. Then the WDW equation reads
\beq
\frac{\hbar^2}{3}\frac{\partial^2\Psi_1}{\partial q_f^2}+(\Lambda q_f-3)\Psi_1=0,
\eeq
where $\Psi_1=\Psi_1(q_f)$. The general solution is given by
\beq
\Psi_1(q_f) =b_1 Ai\left[ \left(\frac{-\sqrt{3} }{\Lambda\hbar}\right)^{\frac{2}{3}} 
\left(3-q_f\Lam\right)\right] 
+ b_2 Bi\left[\left(\frac{-\sqrt{3}}{\Lambda\hbar}\right)^{\frac{2}{3}} 
\left(3-q_f\Lam\right)\right]
\eeq
As before, to match with the $t=1$ part of the path integral 
expression in eq. \ref{eq:Gamp_rbc_int_exact}, we choose the following,
\beq
b_1=-ib_2=\frac{1}{2}e^{-i\pi/3}
\eeq
This fixes the constants $b_1$, $b_2$ uniquely at $t=1$. 
Combining $t=0$ and $t=1$ parts together, we get the total 
wave-function $\Psi(P_i,q_f)=\Psi_0(P_i)\Psi_1(q_f)$. 
Hence, path integral indeed generates the solutions of the WDW equation 
for all values of initial momentum $P_i$ and $\beta$ (alternatively, for a given $P_i$, it is a solution for all $\beta$ and vice-versa.). 
The exponential pre-factor in eq. \ref{eq:Gamp_rbc_int_exact} coming from the topological 
Gauss-Bonnet ($\alpha$) part cannot be captured via the WDW equation. 
However, if we demand that the full wave function, given in 
eq. (\ref{eq:Gamp_rbc_int_exact}) obtained from the path-integral approach, 
keeping the contribution of $\alpha$ to satisfy the WDW equation, 
which is certainly a more stringent criterion, we notice that only 
for $P_i = -3i\sqrt{k}$ the WDW equation is satisfied at the semi-classical level. 
To see this, we will assume $P_i=-iy,\beta=-i\frac{\Lambda}{2}x$, 
which is the relevant range of the parameters to make the total 
wave-function real. Using these relabellings, the WDW equation reads in the semi-classical limit:
\bea
\label{eq:GB_Con}
4\al\left(k+\frac{P_i^2}{9}\right)\left[\frac{\al\Lam x}{3k}
\left(k+\frac{P_i^2}{9}\right)+\left(1-\frac{2 i P_i}{3x\sqrt{k}}
+\frac{1}{x^2}\right)^{1/2}\right]=0.
\eea
Now, assume $P_i \neq \pm 3i\sqrt{k}$, then eq. \ref{eq:GB_Con} 
will be satisfied if the term inside the square bracket vanishes. Since the $\alpha$ dependent term also contains $\Lambda$, upon vanishing, this will make $\alpha$ and $\Lambda$ dependent at arbitrarily chosen $(x,y)$ point.
This certainly doesn't make sense as $\alpha$ and $\Lambda$ are two independent parameters of our theory. Alternatively, if we take $\alpha$ and $\Lambda$ independently, then the eq. \ref{eq:GB_Con} equation is satisfied for a 
 given $y$, only for specific $x$ values, which is not what we wanted. On the other hand, for 
$P_i = \pm 3i\sqrt{k}$, the eq. \ref{eq:GB_Con} satisfied identically for 
all values of $\alpha$, $\Lambda$ and $x$, which makes sense and is along the line of our expectation. From this argument, we conclude allowing Gauss-bonnet 
picks ``only" specific values of $P_i$ which implement no boundary. 
Hence, eliminating all solutions except those no-boundary solutions. 
The last statement supports that no-boundary solutions give the dominant 
contribution to the gravitational path integral at the semi-classical level.
\\

Now, we would like to analyse the no-boundary criterion (vanishing $q$) 
from the canonical perspective, which reads
\begin{equation}
\hat{Q}_i\Psi_0(P_i)=0\,\,\rightarrow\,\,\frac{\partial \Psi_0}{\partial P_i}=0.
\end{equation}
Substituting $\Psi_0(P_i)$ and then taking the derivative, 
we get after simple algebra in the semi-classical limit the following constraint 
\begin{equation}
\label{q0}
\left(y-\frac{3}{x}\right)+\left(9-\frac{6y}{x}+\frac{9}{x^2}\right)^{\frac{1}{2}}=0 \, ,
\end{equation}
where we substituted $P_i=-i y$ and $\beta=-i\frac{\Lambda}{2}x$. 
From eq. \ref{q0}, we get $y=3$ and $0\leq x\leq 1$ for the Hartle-Hawking 
no-boundary Universe agreeing with the previous findings. 
Similarly to the case of the WDW equation, demanding the entire wave function given 
in eq. (\ref{eq:Gamp_rbc_int_exact}), including the Gauss-Bonnet contribution, 
to satisfy the above no-boundary criterion:
\begin{equation}
\hat{Q}_iG_{\rm RBC}=0\,\,\rightarrow\,\,\frac{\partial G_{\rm RBC}}{\partial P_i}=0 \, .
\end{equation}
We obtain the semi-classical constraint as:
\begin{equation}\label{eq:semi_cons_al}
\al\Lam x\left(1-\frac{y^2}{9}\right)+\left(y-\frac{3}{x}\right)
+\left(9-\frac{6y}{x}+\frac{9}{x^2}\right)^{\frac{1}{2}}=0,
\end{equation}
Remarkably, again the solution $y=3$ in the interval $0\leq x \leq 1$ 
identically satisfies the above constraint for all values of $\al$ and $\Lambda$, 
thereby realizing the Hartle-Hawking no-boundary proposal. 
This demonstrates that the conclusions derived from the canonical 
perspective are aligned with those obtained from the path integral approach, 
even in the presence of the Gauss-Bonnet term.

We note that this discussion may well justify the choice of measure 
used in the path integral in previous sections.

\section{Discussion and outlook}
\label{Conc}

In this article, we investigate the gravitational path integral of 
Einstein-Gauss-Bonnet gravity in four spacetime dimensions.
Our study focuses on the Lorentzian path integral within the mini-superspace 
approximation. We extensively examine the effects of imposing the 
Robin boundary condition (fixing the linear combination of scale factor 
and its conjugate momentum) at the initial time, and the Dirichlet 
boundary condition (fixing the scale factor) at the final time and investigate 
the role played by the Gauss-Bonnet sector of gravity

We carefully construct the surface terms needed 
for the Einstein-Gauss-Bonnet gravity with Robin boundary condition 
at an initial time and Dirichlet at the final, and supplement to the 
action to ensure a consistent variational problem. The path-integral is then studied
with this total action.

Utilizing the relations connecting the Robin boundary condition (RBC) path integral as 
an integral transform of the Neumann BC path-integral, 
and drawing upon the results of past studies \cite{Narain:2022msz}, 
where the exact expression for the NBC path-integral has been derived, 
we can obtain the exact expression for the RBC path integral. 
This result is a product of two Airy functions, where one is a function 
of the initial parameters $P_i$ \& $\bt$, and the other is a function of 
the final parameter $q_f$, with an exponential pre-factor that also 
depends on the initial parameters alone. An important observation 
regarding the path-integral is that the term dependent on the 
Gauss-Bonnet coupling ($\al$) appears only in the exponential pre-factor.
When put together, all this gives the exact transition amplitude 
in the RBC case, which is given in eq. (\ref{eq:Gamp_rbc_int_exact}). 
In the limit $\bt\to0$ this correctly reproduces the exact
NBC transition amplitude, which is given in eq. (\ref{eq:Gbd0bd1_full_nbc}). 
This exact transition amplitude for the RBC 
including the Gauss-Bonnet effects, was first derived in \cite{Ailiga:2023wzl}.

We then examined the nature of transition amplitude in the limit
$\hbar\to 0$ (also known as the semi-classical limit). This allowed us 
to identify and focus on configurations that contribute dominantly 
to the path-integral. Primarily, the Gauss-Bonnet 
contribution, which appears as an exponential pre-factor only
shows that in the $\hbar\to0$ limit, two configurations 
give dominant contribution: $P_i = -3i\sqrt{k}$ 
and $P_i = 3i\sqrt{k}$. The former corresponds to 
a stable configuration, while the later is unstable. Interestingly, 
$P_i = -3i\sqrt{k}$ coincides with the configuration for the 
Hartle-Hawking no-boundary Universe, where perturbations are well-behaved. 
In a sense, the Gauss-Bonnet
contribution naturally picks and favours the Hartle-Hawking no-boundary condition.  

Later, we investigated the $\hbar\to0$ limit of other terms in the 
RBC transition amplitude for the case $P_i = -3i\sqrt{k}$ 
(Hartle-Hawking no-boundary), a choice deeply motivated by the 
stability of perturbations associated with this configuration. 
This analysis revealed that $\beta = \beta_{\rm dom} = -i \Lambda / 2\sqrt{k}$ 
yields the dominant contribution. Upon rescaling $\beta = \beta_{\rm dom} x$, 
we observed that for $x \leq 1$, we obtain the same Hartle-Hawking 
exponential pre-factor $e^{6k^{3/2}/\hbar \bar{\Lambda}}$, 
where $\bar{\Lambda}$ is given in eq. (\ref{eq:HHrbcFactor}). 
For $x>1$, the argument of the exponential pre-factor monotonically 
decreases from positive weighting to negative weighting, asymptotically 
approaching $e^{-\frac{6k^{3/2}}{\hbar\Lambda}\left(1-\frac{4\alpha\Lambda}{3}\right)}$. 
This analysis shows that the allowed region where we 
correctly reproduce the Hartle-Hawking exponential 
prefactor is $0\leq x \leq 1$.

We then proceeded to study the saddle point analysis of the 
$\tilde{p}$ integral \ref{eq:Bint_def_Bact}. Making use of the 
Picard-Lefschetz technique, we evaluated the oscillatory integral and 
obtained the amplitude expression in the saddle point approximation, 
which indeed matches with the semiclassical results of exact transition amplitude.
The saddle analysis of the $\tilde{p}$-integration reveals two 
saddle points in the complex $\tilde{p}$ plane, both lying on the imaginary axis. 
The saddle point $\tilde{p}_1$ remains fixed, independent of $x$, 
with a vanishing initial geometry, while the saddle point $\tilde{p}_2$ 
varies with respect to $x$, representing a non-vanishing initial geometry, 
and tends towards infinity (on the negative imaginary axis) as 
$x$ approaches 0 (Neumann limit). For all values of $x$, only one of the 
two saddle points is relevant. For $0 < x < 1$, the saddle point $\tilde{p}_1$ 
is relevant. In this regime, we obtain the correct Hartle-Hawking 
exponential pre-factor $e^{6 k^{3/2}/\hbar \bar{\Lam}}$. For $x > 1$, 
the saddle $\tilde{p}_2$ becomes relevant. At $x = 1$, there is a degenerate 
situation where $\tilde{p}_1=\tilde{p}_2$ (both relevant), resulting in 
$e^{6 k^{3/2}/\hbar \bar{\Lam}}$ (Hartle-Hawking) with additional 
numerical pre-factors. Thus, the range of $x$ for which only the saddle 
point $\tilde{p}_1$ is relevant and produces the exact exponential 
pre-factor $e^{6 k^{3/2}/\hbar \bar{\Lam}}$ of the Hartle-Hawking 
no-boundary Universe is $0 < x < 1$.

In the subsequent part of the article, we explored the relationship 
between the results obtained through the path integral approach and 
those derived from the canonical approach. Although, the topological 
Gauss-Bonnet ($\alpha$) part cannot be captured via the 
Wheeler-DeWitt (WDW) equation. However, when demanding that the 
full wave function, as given in eq. (\ref{eq:Gamp_rbc_int_exact}), 
satisfies the WDW equation, it led to the conclusion that this condition 
is only satisfied for the values $P_i=\pm 3i\sqrt{k}$ for all values of 
$\alpha$. Also, when demanding the entire wave function to satisfy the 
no-boundary criterion from a canonical perspective, we find that \( P_i = -3i\sqrt{k} \) 
in the limit \( 0 \leq x \leq 1 \) is the configuration for which it is satisfied for 
all values of \( \alpha \). Remarkably, these findings align with the results 
obtained from the path integral approach, where \( P_{i} = \pm 3i\sqrt{k} \) 
were identified as the dominant contributions in the semiclassical limit for 
all values of \( \alpha \). Moreover, \( P_i = -3i\sqrt{k} \) and \( 0 \leq x \leq 1 \) 
is the configuration for which the Hartle-Hawking no-boundary proposal is realized.

Our investigations show the important non-trivial role played by the 
Gauss-Bonnet sector of the gravitational action in favouring 
the initial configurations which lead to 
the Hartle-Hawking no-boundary Universe. 
Gauss-Bonnet term arises in the low 
Energy-effective action of the heterotic string theory 
\cite{,Cheung:2016wjt,Zwiebach:1985uq,Gross:1986mw,Metsaev:1987zx}
with $\al>0$. Although it is topological in nature in four spacetime
dimensions and is expected to not play any role in the 
dynamical evolution, but our analysis clearly shows
its contribution to the path integral. When Gauss-Bonnet term
is not ignored in path-integral studies, then for $\al>0$
(which is also the same sign appearing in low
energy effective action of the heterotic string theory)
it naturally picks and favors initial configurations which 
correspond to Hartle-Hawking no-boundary Universe. 
Moreover, Gauss-Bonnet modifications 
won't alter the stability analysis done for Einstein-Hilbert 
gravity \cite{DiTucci:2019dji}. This is because Gauss-Bonnet in 
four spacetime dimension is topological. Perturbative 
analysis like the one done in \cite{DiTucci:2019dji} (see also \cite{Ailiga:2024nkz})
is for small perturbations which are expected to not 
change the background topology,
and hence will be unaffected by the Gauss-Bonnet gravity. 

It has been shown in the literature \cite{Jonas:2020pos} that the 
no-boundary proposal remains robust even with the inclusion of 
higher-order terms in the Riemann curvature. These higher-order 
terms are expected to arise due to quantum gravity corrections. 
Our investigation into the Einstein-Gauss-Bonnet theory 
incorporates such higher-order terms, and the subsequent results align with this notion.


\bigskip
\centerline{\bf Acknowledgements} 
We are thankful to the organizers of the international conference QG@RRI 2023 
for hosting the wonderful event and giving us a chance to present our work in it. 
We appreciate the warm hospitality and the vibrant platform the conference 
provided for exciting discussions. We also thank the anonymous referees for their helpful suggestions.


\appendix
\section{Computation of Fluctuation Prefactor}
\label{Append A}
In this appendix, we detail the computation of the fluctuation prefactor for all the boundary conditions discussed in the main text, following \cite{Ailiga:2023wzl,doi:10.1142/7305}. Consider a one-dimensional quantum mechanical system described by the action
\begin{equation}\label{Ap:action}
    S[q(t)]=\int_{0}^{1} \, \frac{m\dot{q}^2}{2}-V(q)\, dt.
\end{equation}
The path integral we wish to compute is when the Dirichlet boundary condition is imposed on both boundaries (DBC–DBC). The path integral is 
\begin{equation}\label{Ap:G_DD_def}
    \bar{G}[q_f;q_i]_{DD}=\int_{q(0)=q_i}^{q(1)=q_f}\mathcal{D}q(t)e^{iS[q(t)]}.
\end{equation}
The above path integral can be computed exactly for a quadratic potential of the form $V(q)=A q^2+B q+C$, where,  $A, B$, and $C$ are $q$-independent constants. For this potential, the exact path integral acquires the form \cite{doi:10.1142/7305}
\begin{equation}\label{Ap:G_DD_VV}
    \bar{G}[q_f;q_i]_{DD}=\mathcal{F}_{DD} e^{iS[q_{cl}(t)]}
\end{equation}
where,
\begin{equation}\label{Ap:VVPMD}
    \mathcal{F}_{DD}=\left[-\frac{1}{2\pi i\hbar}\frac{\partial^2 S[q_{cl}(t)]}{\partial q_i\partial q_f}\right]^{1/2}
\end{equation}
is the Van Vleck-Pauli-Morette fluctuation determinant for the DBC-DBC case, and $S[q_{cl}(t)]$, is the on-shell action corresponding to the classical solution $q_{cl}(t)$, that satisfies the on-shell equation
\begin{equation}\label{Ap:EOM}
    m\ddot{q}_{cl}=-V''(q_{cl})
\end{equation}
subject to the Dirichlet-Dirichlet boundary condition. \\

Utilizing the above result, we proceed to compute the path integral with Neumann boundary condition (NBC) on the initial hypersurface $p(0) = p_i$, and Dirichlet boundary condition on the final hypersurface, $q(1) = q_f$. The path integral we wish to evaluate is
\begin{equation}
    \bar{G}[p_i;q_f]_{ND}=\int_{p(0)=p_i}^{q(1)=q_f}\mathcal{D}q(t)e^{iS[q(t)]}
\end{equation}
where, $p(t)$ is the canonical momentum associated to $q(t)$. The above integral again can be evaluated exactly using the following relation \cite{Ailiga:2023wzl,doi:10.1142/7305},
\begin{equation}\label{n1}
    \bar{G}[p_i;q_f]_{ND}=\int_{-\infty}^{+\infty}dq_i \, e^{ip_iq_i/\hbar}\times\mathcal{F}_{DD} e^{iS[q_{cl}(t)]}
\end{equation}
Extending the above definitions and relations to the minisuperspace quantum cosmology, eq. (\ref{eq:frwmet_changed}), and using the relations in eq. (\ref{eq:subsNBC}), the expression for the DBC–DBC path integral reads \cite{Feldbrugge:2017kzv}
\begin{equation}\label{Ap:G_DD_PI}
\begin{split}
    \bar{G}_{\rm DBC}[q_f, t=1; q_i, t=0]
=\sqrt{\frac{3 i}{4\pi \hbar N_c}}\exp\left[\frac{i}{\hbar}\left\{\frac{N_c^3\Lambda^2}{36} + N_c\left(-\frac{\Lambda}{2}(q_i+q_f) + 3 k\right)\right.\right.\\
\left.\left.+\frac{1}{N_c}\left(-\frac{3}{4}(q_f-q_i)^2\right)\right\}\right],
\end{split}
\end{equation}
where the exponent corresponds to the on-shell action for the DBC-DBC case, and the prefactor is the Van Vleck-Pauli-Morette fluctuation determinant ($\mathcal{F}_{DD}$). Note that the prefactor matches exactly with the expressions obtained in \cite{Ailiga:2023wzl}. Furthermore, applying the relation, eq. (\ref{n1}, the path integral for Neumann - Dirichlet boundary conditions is given by
\begin{equation}\label{Ap:G_ND_PI}
\bar{G}_{\rm NBC}[q_f, t=1; \pi_i, t=0]
= \exp \biggl[
\frac{i}{\hbar}\biggl\{
\frac{\Lam^2 N_c^3}{9}  
- \frac{\Lam \pi_i N_c^2 }{3} 
+ \left(3k - \Lam q_f + \frac{\pi_i^2}{3} \right) N_c
+ q_f \pi_i
\biggr\}
\biggr] \, .
\end{equation}
Here, as in the DBC–DBC case, the prefactor and exponent correspond to the fluctuation determinant and the on-shell action, respectively. It is crucial to note that the prefactor multiplied to the exponent is unity, 
\begin{equation}\label{fn}
    \mathcal{F}_{ND} = 1,
\end{equation}
which is the fluctuation determinant for the Neumann-Dirichlet case. This agrees with the eq. (\ref{eq:GbarNBC_ms}) in the main text and also align with the results obtained in \cite{Ailiga:2023wzl,doi:10.1142/7305}. 

One may extend the above analysis for the most general linear Robin boundary condition (RBC) on the initial hypersurface and the Dirichlet boundary condition (DBC) on the final hypersurface. The path integral we wish to evaluate is 
\begin{equation}\label{Ap:G_RD}
    \bar{G}[q_f;P_i,\beta]_{RD}=\int_{p(0) + \beta q(0)=P_i}^{q(1)=q_f}\mathcal{D}q(t)e^{iS[q(t)]}.
\end{equation}
Again, the above path integral can be done exactly using the following relation \cite{Ailiga:2023wzl},
\begin{equation}\label{Ap:RD_FT_def}
    \bar{G}[q_f;P_i,\beta]_{RD}=\int_{-\infty}^{+\infty}dq_i \, e^{ip_iq_i/\hbar} e^{-i\beta q_i^2/2\hbar}\times\mathcal{F}_{DD} e^{iS[q_{cl}(t)]}.
\end{equation}
Therefore, utilizing the result in eq. (\ref{Ap:G_DD_PI}), the exact expression for the RBC-DBC path integral reads




\begin{equation}\label{r}
\begin{split}
\bar{G}_{\rm RBC}[q_f, t=1; P_i, t=0] = \frac{1}{\sqrt{1+\frac{2}{3}N_c\beta}}
\exp\left[\frac{1}{18(3+ 2 N_c \bt)} \biggl\{
\bt \Lam^2 N_c^4 + 6 \Lam^2 N_c^3\right. \\
+
\left.N_c^2 \{108 \bt  k-18 \Lam  (P_i+\bt q_f)\}
+18 N_c \left\{9
k+P_i^2+ q_f \left(\bt^2 q_f-3 \Lambda \right)\right\}
+ 54 P_i q_f
\biggr\} \right].
\end{split}
\end{equation}
where the quantity multiplied to the exponent is the fluctuation prefactor,
\begin{equation}
    \mathcal{F}_{RD} = \frac{1}{\sqrt{1+\frac{2}{3}N_c\beta}}.
\end{equation} 
In the $\beta\rightarrow 0$ limit (NBC-DBC), it goes to eq. \ref{fn}. 


\end{document}